\documentclass[compsoc, conference, a4paper, 10pt, times]{IEEEtran}

\usepackage{cite}
\usepackage{amsthm}
\usepackage{amsmath,amssymb,amsfonts}
\usepackage{algorithmic}
\usepackage{graphicx}
\usepackage{textcomp}
\usepackage{xcolor}
\usepackage{booktabs}
\usepackage[hidelinks]{hyperref}
\usepackage[ruled, lined, linesnumbered, commentsnumbered, longend]{algorithm2e}
\usepackage[T1]{fontenc}
\usepackage[english]{babel}
\usepackage{graphicx}
\usepackage{amsmath, xparse}
\usepackage{algorithmic}
\usepackage{multicol}
\DeclareGraphicsExtensions{.png,.pdf}
\usepackage{multirow}
\usepackage{xcolor}
\usepackage{caption}
\usepackage{subcaption}
\usepackage{tabularx}
\usepackage{adjustbox}
\usepackage{pgfplots}
\usepackage{pgfplotstable}
\usepackage{tikz}
\usepackage{xcolor}
\usepackage{pgf-pie}
\usepackage{float}

\pgfplotsset{
    compat=1.17,
}

\definecolor{color1}{HTML}{E68193}
\definecolor{color2}{HTML}{AA8F43}
\definecolor{color3}{HTML}{58A141}
\definecolor{color4}{HTML}{459E97}

\definecolor{color5}{HTML}{ff914d}
\definecolor{color6}{HTML}{0097b2}
\definecolor{color7}{HTML}{0cc0df}
\definecolor{color8}{HTML}{7a72bd}
\definecolor{color9}{HTML}{a8879d}
\definecolor{color10}{HTML}{fde9ff}
\definecolor{color11}{HTML}{FFC1E7}
\definecolor{color12}{HTML}{FEBBCC}
\definecolor{color13}{HTML}{FFCCCC}
\definecolor{color14}{HTML}{FFDDCC}
\definecolor{color15}{HTML}{FFEECC}

\definecolor{color16}{HTML}{008170}
\definecolor{color17}{HTML}{005B41}
\definecolor{color18}{HTML}{232D3F}
\definecolor{color19}{HTML}{0F0F0F}

\definecolor{color20}{HTML}{190482}
\definecolor{color21}{HTML}{7752FE}
\definecolor{color22}{HTML}{8E8FFA}
\definecolor{color23}{HTML}{C2FFFF}

\definecolor{colorw2}{HTML}{643843}
\definecolor{colorw3}{HTML}{99627A}
\definecolor{colorw5}{HTML}{C88EA7}
\definecolor{colorw7}{HTML}{E7CBCB}

\definecolor{colorlbu}{HTML}{E68193}
\definecolor{colorlsp}{HTML}{AA8F43}
\definecolor{colorlbd}{HTML}{58A141}
\definecolor{colorlba}{HTML}{459E97}

\definecolor{coloruser100}{HTML}{F1D3B3}
\definecolor{coloruser1000}{HTML}{C7BCA1}
\definecolor{coloruser10000}{HTML}{8B7E74}

\definecolor{colora5}{HTML}{C63D2F}
\definecolor{colora10}{HTML}{E25E3E}
\definecolor{colora15}{HTML}{FF9B50}
\definecolor{colora20}{HTML}{FFBB5C}

\newtheorem{theorem}{Theorem}
\theoremstyle{definition}
\newtheorem{definition}{Definition}[section]

\newcommand{\mysys}{$LDP$\textendash$SmartEnergy$}

\begin{document}

\title{Local Differential Privacy for Smart Meter Data Sharing}

% The author information is skipped here, but can be used to include author information in the publication.
% \iffalse
\author{\IEEEauthorblockN{Yashothara Shanmugarasa}
\IEEEauthorblockA{\textit{University of New South Wales} \\
Sydney, Australia \\
y.shanmugarasa@unsw.edu.au}
\and
\IEEEauthorblockN{M.A.P. Chamikara}
\IEEEauthorblockA{\textit{CSIRO’s Data61} \\
Melbourne, Australia \\
chamikara.arachchige@data61.csiro.au}
\and
\IEEEauthorblockN{Hye-young Paik}
\IEEEauthorblockA{\textit{University of New South Wales} \\
Sydney, Australia \\
h.paik@unsw.edu.au}
\and
\IEEEauthorblockN{Salil S. Kanhere}
\IEEEauthorblockA{\textit{University of New South Wales} \\
Sydney, Australia \\
salil.kanhere@unsw.edu.au}
\and
\IEEEauthorblockN{Liming Zhu}
\IEEEauthorblockA{\textit{CSIRO’s Data61} \\
Sydney, Australia \\
liming.zhu@data61.csiro.au}
%% IEEE format can accommodate up to six authors this way
}
% \fi

\maketitle

\begin{abstract}
Energy disaggregation techniques, which use smart meter data to infer appliance energy usage, can provide consumers and energy companies valuable insights into energy management. However, these techniques also present privacy risks, such as the potential for behavioral profiling. Local differential privacy (LDP) methods provide strong privacy guarantees with high efficiency in addressing privacy concerns.
However, existing LDP methods focus on protecting aggregated energy consumption data rather than individual appliances. Furthermore, these methods do not consider the fact that smart meter data are a form of streaming data, and its processing methods should account for time windows. In this paper, we propose a novel LDP approach (named {\mysys}\footnote{The source code of {\mysys} will be publicly available after acceptance.}) that utilizes \textit{randomized response} techniques with sliding windows to facilitate the sharing of appliance-level energy consumption data \textit{over time} while not revealing individual users' appliance usage patterns.
Our evaluations show that {\mysys} runs efficiently compared to baseline methods. The results also demonstrate that our solution strikes a balance between protecting privacy and maintaining the utility of data for effective analysis.
\end{abstract}


\section{Introduction}
\label{sec:intro}

Smart meters digitally track real-time energy consumption at a particular location (e.g., a house, a building) and transmit the corresponding data to energy companies for analysis~\cite{marks2021differential}.
Smart meters offer various benefits, including accurate energy usage billing and improved grid operation stability \cite{marks2021differential}.

Energy disaggregation (ED) refers to the process of determining the energy usage of individual appliances (e.g., within a building). 
ED helps consumers understand device usage patterns and appliance status to reduce inadvertent faults. ED can assist energy companies in identifying the impact of individual appliances on overall energy consumption, segmenting consumers based on energy usage patterns, and troubleshooting issues in real-time~\cite{iqbal2021critical}.

However, releasing energy consumption information to energy companies or other third parties raises privacy concerns.
For instance, applying ED techniques could enable real-time monitoring of household activities, appliance usage tracking, behavioral profiling (e.g., meal times, exercise patterns), and identity theft \cite{li2010secure,kroger2018unexpected}.

The literature shows various privacy-preserving techniques to reduce the privacy risks posed by smart meter data sharing~\cite{marks2021differential}. 
Approaches leveraging differential privacy (DP) ~\cite{dwork2006calibrating} are preferred over other privacy preservation techniques, such as cryptographic approaches, due to their robust privacy guarantees and high efficiency. Cryptographic approaches can often be computationally complex and suffer from performance degradation~\cite{gai2022efficient}.

DP schemes achieve data protection by introducing calibrated noise to the data. In global differential privacy (GDP), a trusted curator applies noise to the data. In local differential privacy (LDP), the data owners are responsible for randomizing their data before releasing it to third parties. 
It is shown that LDP offers stronger privacy guarantees than GDP at the expense of slightly lower utility (i.e., the accuracy and usefulness of data analysis results) \cite{liu2022gdp}.

However, to apply LDP to disaggregated smart meter data, we need to consider the following two aspects~\cite{hassan2019differential, gai2022efficient}:
\begin{itemize}
    \item Previous approaches were designed to obtain global insights from aggregated energy data, for instance, to calculate the total or average energy consumption. However, these approaches cannot be directly applied to data obtained through ED, as they could reveal individual appliance information, such as the list of appliances and their energy usage, due to insufficient levels of randomization.
    \item Smart meter data are a form of streaming data \cite{smart-meter-data-review2019}, which requires time-window-based processing to prevent the gradual disclosure of sensitive information. Existing LDP methods do not account for this aspect. 
\end{itemize}

In this paper, we introduce a novel LDP protocol, known as {\mysys}, for enhancing the privacy-preserving sharing of disaggregated energy consumption data from household appliances. The novelty of our work is summarised as follows:
\begin{itemize}
\item 
We leverage a unique modeling technique to process disaggregated energy data, transforming them into discrete values via mapping and quantization. Then, we randomize energy consumption data using a randomization mechanism based on randomized response \cite{fox2015randomized} and optimized unary encoding \cite{wang2017locally, mahawaga2022local} to guarantee differential privacy. {\mysys} still prioritizes protecting individual data while generating global insights (i.e., insights from aggregated energy data across all appliances from all homes). This is accomplished by adhering to the principles of DP, ensuring that individual insights from each home's appliances are not discernible. 
\item We utilize a privacy budget to signify the predefined threshold for allowable privacy loss \textit{over time periods}. Recognizing the streaming nature of energy data, we apply privacy protection measures by adopting a sliding window approach to counteract the gradual exposure of data over time.
\end{itemize}

We implement and demonstrate the effectiveness of our solution through a use case where top-k appliance usage patterns and the appliances list are obtained from the energy consumption data of households.  

The rest of this paper is organized as follows. Section \ref{sec:preliminaries} introduces the technical foundation and concepts relevant to {\mysys}. Section \ref{sec:approach} presents the overview of our approach {\mysys}, its architecture, and discusses the key components. Section \ref{sec:reanddisc} comprehensively evaluates the performance of {\mysys} with two datasets and discusses the results. Section \ref{sec:related} provides an overview of related work covering different privacy-preserving techniques for smart meter data. Section \ref{sec:conclusion} concludes the paper and offers insights into potential future directions.

\section{Preliminaries}
\label{sec:preliminaries}

This section introduces the technical foundation and concepts relevant to {\mysys}. These include individual appliance energy consumption modeling (i.e., energy disaggregation over appliances), LDP, Randomized Aggregatable Privacy-Preserving Ordinal Response (RAPPOR), variants of unary encoding, and adaptive budget division methods.

\subsection{Energy Disaggregation over Appliances}

A smart meter measurement, $V_t$ represents the total energy consumption at a particular location (e.g., a house or a building) measured at a time $t$. 
Energy disaggregation refers to identifying the energy consumption of individual appliances contributes to $V_t$. We model this as follows. 
Suppose a location (e.g., a single household) has $n$ appliances. Then $A = \left\{a_1, a_2, \dots, a_n\right\}$, the energy consumption of the appliances $\left\{[a_{1t},v_{1t}], \dots, [a_{nt},v_{nt}]\right\}$ at time $t$ can be represented as denoted in Eq. (\ref{eq:nilm}). The variable $v_{it}$ represents the energy consumption of $a_i$, while $E_t$ represents the error during energy disaggregation.

\begin{equation}
\label{eq:nilm}
\begin{split}
V_t = \sum_{i=1}^{n}((a_{it}\times v_{it}) + E_t)   \quad\quad 
_{\text{\quad \quad \hspace{0.05cm} 0 if appliance $a_i$ is off at time $t$.}}^ {a_{it}  = 1  \hspace{0.1cm} \text{if appliance $a_i$ is on at $t$;}}
\end{split}
\end{equation}

\subsection{Local Differential Privacy (LDP)}
\label{subsec:ldp}

LDP is a privacy preservation technique where data owners can apply randomization or noise to their data before sharing them with a third party.
%a curator (or a third-party analyst). 
If a randomized algorithm $\mathcal{A}$ satisfies Eq. (\ref{eq:LDP}), then it provides $\varepsilon$-local differential privacy. 

\begin{definition}

An algorithm $\mathcal{A}$ satisfies $\varepsilon$-local differential privacy for any user if for any two data values $v_1$ and $v_2$, and for any output $Q$ within the range of outputs of algorithm $\mathcal{A}$ ($Q\subseteq Range(\mathcal{A})$), the following Eq. (\ref{eq:LDP}) holds ($\varepsilon$ is the privacy budget and it represents the degree of privacy) \cite{dwork2016calibrating, erlingsson2014rappor}:

\begin{equation}
\label{eq:LDP}
    Pr\left[ \mathcal{A}(v_{1})\in Q \right]\le e^{\varepsilon}\times\Pr[\mathcal{A}(v_{2}\in Q)]
\end{equation}
where the probability is taken over the random measure implicit in $\mathcal{A}$.
\end{definition}

\subsection{Randomized Aggregatable Privacy-Preserving Ordinal Response (RAPPOR)}

RAPPOR is a technique proposed in \cite{erlingsson2014rappor} to provide strong privacy guarantees for users participating in crowdsourcing data collection. The technique has since been applied in many LDP approaches \cite{ding2017collecting, zheng2020privacy}.
RAPPOR encodes an input data instance (i.e., user responses) into a binary string, $B$, which is randomly perturbed before being sent to a data processing server. A binary string, $B$, is composed of $d-1$ zeros and one $1$ at a specific position, $v$. Randomization preserves privacy by randomizing the 0s and 1s in the binary string $B$, creating a new string, $B\acute{}$. During the randomization process, the true value of an input bit-string $B$ is preserved with a probability of $p$, as in Eq. (\ref{eq:bitrandomization}) where 

$\Delta f$ is the sensitivity of a function $f$, which is the maximum impact that a single individual can exert on a function $f$
In RAPPOR's encoding process, sensitivity ($\Delta f$) % (i.e. maximum change in a function \textit{f}'s result with individual's inclusion or exclusion) 
can be defined as the maximum difference between two adjacent encoded bit strings, ${B}(x_{i})$ and ${B}(x_{i+1})$. This difference is limited to two bits.

\begin{equation}
\label{eq:bitrandomization}
   p = \frac{e^{\frac{\varepsilon}{\Delta f}}}{1+e^{\frac{\varepsilon}{\Delta f}}} = \frac{e^{\frac{\varepsilon}{2}}}{1+ e^{\frac{\varepsilon}{2}}}
\end{equation}

\subsection{Unary Encoding and Optimized Unary Encoding}
\label{subsec: OUE}

RAPPOR employs unary encoding (UE) \cite{wang2017locally} to transform the input instance $y_{i}$ into a binary vector $B$ consisting of $d$ number of bits. UE sets a specific bit at position $v$ in $B$ to 1, and all others to 0, then perturb a bit, $B[i]$, into a perturbed bit $B\acute{}$[i] according to Eq. (\ref{eq:UE}).

\begin{equation}
\label{eq:UE}
\Pr[B'[i]=1]=\begin{cases}
        p & \text{if } B[i]=1 \\
        q & \text{if } B[i]=0
\end{cases}
\end{equation}

UE provides $\varepsilon$-LDP~\cite{erlingsson2014rappor}. Specifically, for any bit positions $v_1$ and $v_2$ (of the encoded inputs $x_1$ and $x_2$, respectively) in a d-bit vector, and an output $B$ with sensitivity = 2, UE ensures $\varepsilon$-LDP when the values of $p$ and $q$ are given by $p = \frac{e^{\varepsilon/2}}{1+e^{\varepsilon/2}}$ and $q = \frac{1}{1+e^{\varepsilon/2}}$. The highest value of this ratio occurs when $v_1$ is 1 and $v_2$ is 0, denoted by Eq. (\ref{eq:UE2}).

\begin{proof}
\label{proof:UE}
\begin{equation}  
\label{eq:UE2}
%Let $\varepsilon$ be the privacy budget, and $\Delta f$ be the sensitivity.
% $p =  \frac{1}{(1 + \alpha)} $
% $q =  \frac{1}{(1 + \alpha e^{\frac{\varepsilon}{\Delta f/2}})}$ 
\begin{aligned}
% p &=  \frac{e^{\varepsilon/2}}{1+e^{\varepsilon/2}}, \quad \quad q =  \frac{1}{1+e^{\varepsilon/2}}\\ 
\frac{\Pr[B|v_{1}]}{\Pr[B|v_{2}]} &= \frac{\prod_{i\varepsilon[d]}^{}\Pr[B[i]|v_{1}]}{\prod_{i\varepsilon[d]}^{}\Pr[B[i]|v_{2}]} \\
&\le\left( \frac{\Pr[B[v_{1}] = 1|v_{1}]\Pr[B[v_{2}] = 0|v_{1}]}{\Pr[B[v_{1}] = 1|v_{2}]\Pr[B[v_{2}] = 0|v_{2}]} \right )\\
&= \frac{p}{q} \times \frac{1-q}{1-p}
=e^{\varepsilon}
\end{aligned}
\end{equation}
\end{proof}

The Optimized Unary Encoding (OUE) method improves upon UE by perturbing 0s and 1s differently. There are more 0s than 1s (usually only one 1) in a long binary vector, $B$. OUE reduces the chance of perturbing 0 to 1 ($p_{0\to 1}$) by allocating more budget to transmit 0 bits in their original state. 
By setting $p=1/2$ and $q=\frac{1}{1+e^\epsilon}$\cite{wang2017locally}, and using Equation (\ref{eq:UE2}), OUE guarantees $\epsilon$-LDP for values with sensitivity equal to 2\cite{wang2016using}.

\subsection{Adaptive Privacy Budget over Time}
\label{subsec:adaptive}

The authors in  \cite{10.1145/3514221.3526190} proposed adaptive privacy budget division methods as part of the LDP-IDS (LDP for Infinite Data Stream) framework, which works with streaming data. The traditional approach of concealing a single event in data streams is inadequate for safeguarding user privacy, and the method designed to protect a user's presence in infinite streams is impractical in real-world scenarios. A concept called ``$w$-event privacy'' aims to ensure $\epsilon$-differential privacy ($\epsilon$-DP) for any time window comprising $w$ consecutive time instances, making it a more practical approach for streaming data.
A mechanism that satisfies $w$-event privacy can provide $\varepsilon$-LDP guarantees in any sliding window of size $w$ \cite{wang2019locally, 10.1145/3514221.3526190}.

\subsubsection{Naive methods}
\label{subsec:naive}
There are various adaptive budget division methods for $w$-event privacy. Two naive methods for dividing the privacy budget among timestamps within a window to achieve $w$-event privacy are the \textit{LDP Budget Uniform Method (LBU)} and the \textit{LDP Sampling Method (LSP)}\cite{10.1145/3514221.3526190}. LBU evenly divides the budget among all timestamps, but it may not be efficient for larger windows. The LSP allocates the entire privacy budget $\varepsilon$ to a single timestamp (sampling timestamp) within a window of size $w$. The budget is not used for the remaining $(w-1)$ timestamps, which are instead approximated with the perturbed value from the sampling timestamp. However, the LSP method cannot accurately track changing patterns in data streams, potentially leading to errors.

\subsubsection{Local Budget Distribution (LBD)}
\label{subsec:lbd}

LBD \cite{10.1145/3514221.3526190} is a method that was initially proposed for global DP in \cite{kellaris2014differentially}. 
In contrast to the naive methods (i.e., LBU and LSP), 
LBD uses statistical dissimilarity metrics to help identify the best approach for either approximating current statistics using past data or adding noise to current statistics when publishing them.

The LBD approach consists of three components: private dissimilarity calculation, private strategy determination, and privacy budget allocation \cite{10.1145/3514221.3526190}\footnote{Please see Algorithm \ref{algo:lbdlba} in Appendix \ref{algorithm:adaptive}.}. 
All calculations occur on the user side and utilize true data to derive intermediate steps.
In the context below, $\varepsilon$ represents the privacy budget, $w$ denotes the window size, and $\varepsilon_{rm}$ is the remaining publication budget at the current timestamp. 

In private dissimilarity calculation, a dissimilarity value \textit{dis} is determined by comparing the current perturbed value $ \bar c_t$ (with a fixed dissimilarity budget $\varepsilon_{t,1} = \varepsilon/(2 \times w)$) and last release using Mean Absolute Error (MAE). 
% (refer to Algorithm \ref{algo:lbdlba} Sub Mechanism $M{t,1}$).
In the private strategy determination component, a budget $\varepsilon_{t,2}$ is allocated for the possibility of publishing perturbed data, $\bar{c}_{t,2}$, with a potential publication budget  $\varepsilon_{t,2} = \varepsilon_{rm}/2$). % (refer to Algorithm \ref{algo:lbdlba} Sub Mechanism $M{t,2}$).
This budget will be determined in the privacy budget allocation component\footnote{More details on potential publication budget determination are discussed in the publication budget allocation component and Algorithm \ref{algo:lbdlba}, Sub Mechanism $M_{t,2}$.}.
The private strategy determination algorithm employs a decision-making process that utilizes a comparison of dissimilarity (\textit{dis}) and potential publication error (\textit{err}) to determine whether to utilize the approximation strategy (publish the last released value directly without using up the publication budget $\varepsilon_{t,2}$) or to publish the current data with perturbation using $\varepsilon_{t,2}$. % (refer to Algorithm \ref{algo:lbdlba} Sub Mechanism $M{t,3}$). 

In the publication budget allocation component, the LBD approach distributes the entire budget, $\varepsilon$, in a time window evenly between two parts: private dissimilarity estimation and private strategy determination, with dissimilarity and publication budget being $\varepsilon/2$ each ($\sum_{i=t-w+1}^{t}\varepsilon_{i,1} = \sum_{i=t-w+1}^{t}\varepsilon_{i,2} = \varepsilon/2$). Further, the dissimilarity budget is distributed equally among all timestamps within the window, with each timestamp being allocated a budget of $\varepsilon_{i,1} = \varepsilon/(2\times w)$.

The publication budget $\varepsilon_{i,2}$ is determined using a cost-effective privacy budget allocation strategy, where it is distributed in an exponentially decreasing manner among timestamps. 
As some publications are skipped via approximation, the remaining privacy budget ($\varepsilon_{rm}$) at each current timestamp is calculated by subtracting the dissimilarity and publication budget spent in a window from the total privacy budget. % (refer to Algorithm \ref{algo:lbdlba} Sub Mechanism $M{t,2}$, Line \ref{step:m2}).
Next, half of the remaining budget, $\varepsilon_{rm}$, is allocated as the publication budget, $\varepsilon_{i,2}$, and potential publication error (\textit{err}) is calculated. If the current data is chosen to be published, publication budget $\varepsilon_{i,2}$ will be used in the current timestamp; otherwise, it will be set to 0 and used in the future. 
% \begin{equation}
% \label{eq:errorRemain}
% \varepsilon_{rm} = \varepsilon/2 - \sum_{i=t-w+1}^{t-1}\varepsilon_{i,2}
% \end{equation}

\subsubsection{Local Budget Absorption (LBA)}
\label{subsec:lba}

LBD and LBA have similar components for calculating private dissimilarity and determining private strategy. However, their methods for dynamically allocating the publication budget of $\varepsilon_{i,2}$ over the stream differ\footnote{Algorithm \ref{algo:lbdlba} (refer to Appendix \ref{algorithm:adaptive}) gives the details.}. The LBA approach starts with uniformly distributing the publication budget across all timestamps. If the current data is not published at a particular timestamp, the budget allocated to that timestamp becomes available for future publication. However, when the current data is published at a specific timestamp, the published data absorb all the available budget accumulated from previous skipped publications.
%, using it to publish current statistics with increased utility 
This is to publish current statistics with increased utility~\cite{kellaris2014differentially, 10.1145/3514221.3526190}. 
However, whenever the publication budget is absorbed for the current publication from previous timestamps, an equivalent amount of budget must also be nullified from the immediately succeeding timestamps,
%(refer to Algorithm \ref{algo:lbdlba} Sub Mechanism $M{t,2}$, Lines \ref{step:lbam21}- \ref{step:null}), 
resulting in their publishing outputs becoming null. Nullifying the budget helps prevent exceeding the maximum budget $\varepsilon$. The determination of the publication budget $\varepsilon_{i,2}$ is based on the decision to absorb or nullify, and once decided, it is used to calculate the potential publication error (\textit{err}).
After calculating potential publication error (\textit{err}) with $\varepsilon_{i,2}$, the optimal strategy between publication and approximation is chosen by comparing with dissimilarity \textit{dis}, similar to the LBD mechanism.

The LBD / LBA methods have been shown to provide $\varepsilon$-LDP in \cite{10.1145/3514221.3526190}. Algorithm \ref{algo:lbdlba} presents the pseudo-code for the LBD and LBA process, while Table \ref{tab:TableOfNotations} (refer to Appendix \ref{algorithm:adaptive}) provides the notations and their corresponding descriptions used in Algorithm \ref{algo:lbdlba}.

\subsection{Composition and Post-processing Invariance}
\label{subsec:composition}

The adaptive privacy budget methodologies discussed above are made possible by the properties of differential privacy, such as composition and post-processing invariance.
Composition is the loss of privacy when multiple DP algorithms are applied to the same or overlapping datasets \cite{bun2016concentrated}. When two DP algorithms, $\varepsilon_1$-DP and $\varepsilon_2$-DP are applied to the same or overlapping data sets, the output is ($\varepsilon_1$ + $\varepsilon_2$)-DP \cite{bun2016concentrated}. Parallel composition enables multiple users (or models) to participate in the $\varepsilon$-LDP mechanism.
In parallel composition, when a group of DP algorithms ($M_1, M_2, \dots, M_n$) are applied to disjoint subsets of a dataset ($D_1, D_2, \dots , D_n$), the entire process will provide max \{$\varepsilon_1, \varepsilon_2, \dots , \varepsilon_n$\}-DP for the entire dataset, with each algorithm ($M_i$) providing $\varepsilon_i$-DP for its corresponding subset ($D_i$).
% \cite{zhao2019differential}.
The post-processing invariance or robustness property of DP states that any additional computations on the outputs of a DP algorithm (that are independent of the original database) do not diminish its privacy guarantees.

\section{Methodology}
\label{sec:approach}

\begin{figure*}[ht]
\centering
  \includegraphics[width=0.9\linewidth]{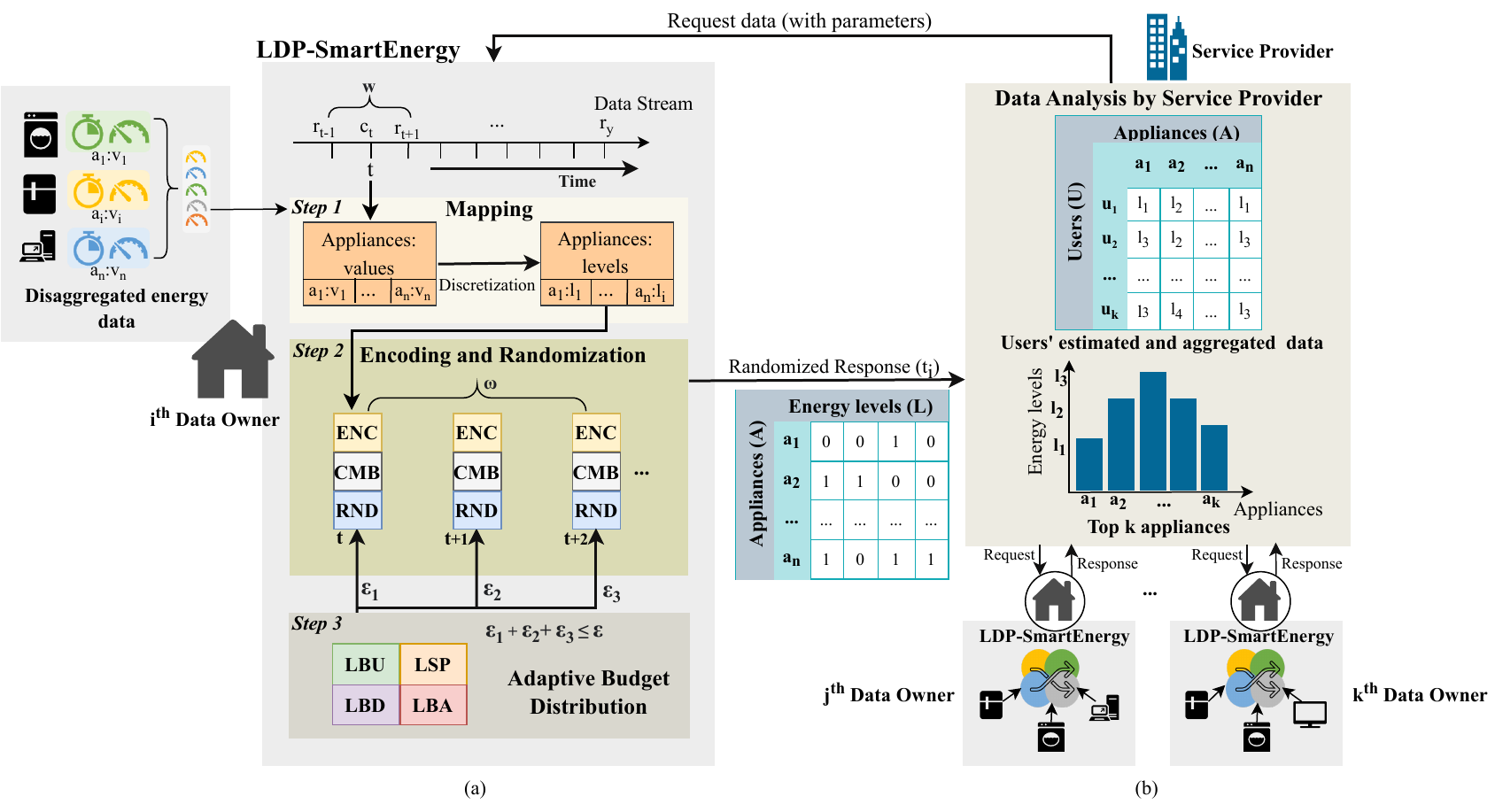}
  \caption{The architecture of {\mysys} (a): On the user (data owner) side, (b): On the service provider side. $\{r_{t-1}, c_t, r_{t+1}, \dots, r_y\}$: Energy consumption data stream, $c_t$: Data of the current timestamp, $\{a_1,v_1, $\dots$, a_n, v_n \}$: appliances with their corresponding energy values, $\{a_1,l_1, $\dots$, a_n, l_i \}$: appliances with their corresponding mapped energy levels, \textbf{ENC}: Encoding, \textbf{CMB}: Combining all appliances encoded data, \textbf{RND}: Randomization, \textbf{$\varepsilon_1, \varepsilon_2, \varepsilon_3, \varepsilon$}: Privacy budgets, \textbf{LBD}: Local budget distribution, \textbf{LBA}: Local budget absorption}
  \label{fig:workflow}
\end{figure*}

An energy consumption data-sharing framework that utilizes {\mysys} is shown in Figure \ref{fig:workflow}. Figure~\ref{fig:workflow}a depicts the processes within {\mysys}, which is implemented and run on the user side. First, {\mysys} takes disaggregated energy consumption data from appliances and applies the following three steps: (1) Mapping the individual appliances' energy consumption to a given range of energy levels (step 1 in Figure \ref{fig:workflow}a), (2) Encoding energy consumption levels, generating 1-D vector 
%creating a flattened 1-D vector (i.e., combining all appliance encoded data to one long binary vector) 
with all appliance details, and adding randomization to enforce DP (step 2 in Figure \ref{fig:workflow}a), and (3) Maintaining the budget composition of a data stream using an adaptive budget division methods (step 3 in Figure \ref{fig:workflow}a). Step 3 determines the privacy budget for randomization in each timestamp of step 2.

\subsection{The Rationale}
Unlike existing perturbation techniques (that utilize Laplace, Gaussian,  and Gamma mechanisms) in the literature that add noise to handle numerical energy values, our approach utilizes a mechanism that employs randomized response for better utility and privacy.
Since we handle disaggregated individual appliances, existing methods could potentially reveal individual appliance data (i.e., the appliance list and each appliance's energy value) due to inadequate levels of noise or randomization focusing on protecting aggregated energy consumption data rather than individual appliances. 
\textbf{\mysys} offers a flexible way to ensure the privacy of individual appliances, without revealing the list of appliances. This is achieved through bit-wise randomization. 
\textbf{\mysys} combines all appliance energy data and uses bit-wise randomization to effectively conceal both the list of appliances and the specific energy consumption of individual homes. This is achieved by utilizing a predefined list of appliances that is universally applicable to any household.
Our approach mandates reporting energy consumption exclusively for this predetermined list, regardless of the actual presence of the appliances. 
Consequently, this approach further restricts an attacker from accurately guessing the number of active appliances within a household.
To achieve this task, \textbf{\mysys} introduces a novel preprocessing step that discretizes the continuous energy values into distinct levels before the randomization through a novel protocol (\textbf{\mysys}) based on randomized response. This innovative approach paves the way for enhanced utility with strong privacy guarantees due to randomized response for a problem where randomized response has not traditionally been applied. 

\vspace{-3mm}

\begin{figure}[ht]
    \centering
    \includegraphics[width=0.36\textwidth]{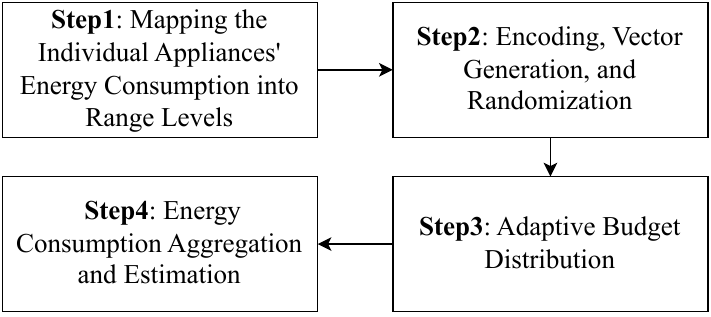}
    \caption{The primary steps of {\mysys}}
    \label{fig:workflow_steps}
\end{figure}

\setlength{\textfloatsep}{0.5cm}% Remove \textfloatsep
\begin{algorithm*}[!h]
\caption{{\mysys} user-side algorithm}
\label{alg1}
\begin{multicols}{2}
\SetKwFunction{LDPTopK}{LDPTopK}
    \SetKwInOut{KwIn}{Input}
    \SetKwInOut{KwOut}{Output}

    \KwIn{\{<$a_1$,$v_1$>, $\dots$, <$a_n$, $v_n$>\} \quad \quad  $\leftarrow$ $n$ number of appliances \& corresponding energy values  \newline
   \{(0,$r_1$):$l_1$, $\dots$, ($r_d$, max):$l_d$\} \quad \quad $\leftarrow$ energy ranges \& the corresponding levels $L$\newline  
    $\varepsilon$ \quad $\leftarrow$ total privacy budget\newline 
    $w$ \quad  $\leftarrow$ window size \newline
    }
    % \KwOut{\{<$a_1$,($l_i$, $\dots$,$l_k$)>, $\dots$, <$a_n$, ($l_i$, $\dots$, $l_k$)>\} \quad $\leftarrow$ DP appliance list with their energy consumption levels over a time}
    \KwOut{\{$B_1$, $\dots$, $B_i$\} \quad $\leftarrow$ 1-D vector of combined and randomized energy consumption over $i$ time periods for all appliances}

    \textbf{Step I: Mapping the Individual Appliances' Energy Consumption into Range Levels} (refer to Section \ref{subsec:map})

    \For {each timestamp $t$} {
     \For {each <$a_i$,$v_i$>} {
          \For {each ($r_d$, $r_i$):$l_i$} {
             \uIf{$v_i > r_d$ and $v_i \le r_i$ }{
                \Return <$a_i$,($l_i$)>
             }     
     }
     }
     \Return \{<$a_1$,$l_1$>, $\dots$, <$a_n$, $l_i$>\} 
     }

    \textbf{Step II: Encoding, Vector Generation and Randomization}

     Declare, $m$ = $(d\times n)$, 1-D binary vector $B$ length of $m$

    \For {each timestamp $t$} {   

    Encode \{<$a_1$,$l_1$>, $\dots$, <$a_n$, $l_i$>\} to $d$ bits of binary vector $v$ (refer to Section \ref{subsec:binaryenc})
    
    % \{$d_1$, $\dots$, $d_n$\} = Each appliance's encoded value of $l_i$ (consisting of $d$ bits of binary vector) (refer to Section \ref{subsec:binaryenc});
    \For {each appliance vector $v$} {
        Append $v$ to $B$
        
        }
    \Return $B$
    % Generate, a 1-D binary vector $B$ of length $m$ by merging the encoded vectors of all appliances
    
    Calculate $\varepsilon$ and randomize $B$ into $RND$ using approaches from part III % (steps \ref{ap1}, \ref{ap2}, or \ref{ap3})

    }
    
   \textbf{Step III: Adaptive Budget Division} \label{step:adaptive}
   
    \For {each timestamp t} {
    
   \textit{Approach 1: LDP budget uniform (LBU)} \label{ap1}
    
   $\varepsilon_1$ = $\varepsilon/w$
   
   \Return RND = randomize $B$ with $\varepsilon_1$

   \textit{Approach 2: LDP sampling method(LSP)} \label{ap2}
   
    \uIf{$t$ == $sampling timestamp$ }{

        $\varepsilon_2$ = $\varepsilon$
    }
    \Else{
        $\varepsilon_2$ = 0
    }

    \Return RND = randomize $B$ with $\varepsilon_2$

    \textit{Approach 3 \& 4: Local budget distribution or Local budget absorption } \label{ap3}    

    % \tcp{Sub Mechanism $M_{t,1}$}
    \tcp{\footnotesize Sub Mechanism $M_{t,1}$}
    
    refer to Algorithm \ref{algo:lbdlba}, lines \ref{step:e1} to \ref{step:discal}

    \tcp{\footnotesize Sub Mechanism $M_{t,2}$}

    refer to Algorithm \ref{algo:lbdlba}, lines \ref{step:m2} to \ref{step:m2error} for LBD and \ref{step:lbam21} to \ref{step:calerr2} for LBA

    \tcp{\footnotesize Sub Mechanism $M_{t,3}$}

    refer to Algorithm \ref{algo:lbdlba}, lines \ref{step:check1} to \ref{step:check3} for LBD and \ref{step:check12} to \ref{step:check22} for LBA
    }
\end{multicols}
\end{algorithm*}

\begin{algorithm}[ht]
\caption{{\mysys} server-side algorithm}
\label{alg3}
\SetKwFunction{LDPTopK}{LDPTopK}
    \SetKwInOut{KwIn}{Input}
    \SetKwInOut{KwOut}{Output}

    \KwIn{\{<$u_1$,$B_1$>, \dots, <$u_k$,$B_k$>\}\quad \quad  $\leftarrow$ $k$ number of users \& corresponding combined 1-D flattened vector with energy levels of given appliances list \newline
    $\varepsilon$ \quad $\leftarrow$ maximum total privacy budget\newline 
    $n$ \quad $\leftarrow$ number of appliances \newline 
    $d$ \quad $\leftarrow$ number of energy levels\newline 
    $sensitivity$ = 2 \quad $\leftarrow$ each appliance sensitivity 
    }
    % \KwOut{\{<$a_1$,$l_1$>, $\dots$, <$a_n$,$l_i$>)>\} \quad $\leftarrow$ Aggregated and estimated energy levels of the appliance list}
    \KwOut{\{<$a_1$,$l_i$,$c_i$>, $\dots$, <$a_1$, $l_i$, $c_j$>, $\dots$, <$a_n$,$l_i$,$c_i$>, $\dots$, <$a_n$, $l_i$, $c_j$>\} \quad $\leftarrow$ Aggregated and estimated energy levels of the appliance list}

  \{ <$a_1$,$p_1$>, $\dots$, <$a_n$,$p_n$>\} = Decode the $B$ into appliance-wise energy levels by dividing $B$ by $d$
  
    \For {each <$a_i$,$p_i$>} {    
     \For {each user $u$} {
     
         Define $p$ = $0.5$, $q$ =  $\frac{1}{(1 + e^{\frac{\varepsilon}{sensitivity/2}})}$ (the probabilities of randomizing a bit using OUE)

        \tcp{\footnotesize Aggregate energy levels of appliances from all users and estimate the corresponding energy level values (steps \ref{ste:ser:sum} - \ref{ste:ser:estmate2})}
        
         $sums$ = sum $p_i$ of each $u$ \label{ste:ser:sum}

         \For {each $sums$ $y$} { \label{ste:ser:estimate}
            \Return $aggregatedPerturbedValues$ = $(y - d * q) / (p - q)$ \label{ste:ser:estmate2}                
        }  
     }
     \Return \{<$a_i$,$l_i$,$c_i$>, $\dots$, <$a_i$, $l_i$, $c_j$>\}  \tcp{\footnotesize $c_i$, $c_j$ estimated count of users that belong to a particular energy level of an appliance after perturbation}
     }
     \Return \{\{<$a_1$,$l_i$,$c_i$>, $\dots$, <$a_1$, $l_i$, $c_j$>\}, $\dots$, \{<$a_n$,$l_i$,$c_i$>, $\dots$, <$a_n$, $l_i$, $c_j$>\}\} 
\end{algorithm}

\subsection{\textbf{\mysys} Overview}

In \textbf{\mysys}, a service provider, as shown in Figure \ref{fig:workflow}b, initiates the process by determining parameters such as the appliance list, privacy budget ($\varepsilon$), data collection time interval, energy level ranges, and window size. The hyperparameters are data-independent and predetermined by service providers based on prior knowledge \cite{dochy1994prior}.

Figure \ref{fig:workflow_steps} shows the primary steps of \textbf{\mysys}. The process in user side starts with mapping to limit the disclosure of energy values of individual appliances while also enabling high efficiency with concurrency. Next, the energy levels are converted into binary vectors for randomization with high privacy and high utility. 

Finally, the service provider concludes a data collection round by aggregating the randomized data from all users.
To demonstrate the high utility of randomized data following the application of {\mysys}, we consider a practical use case where a household shares smart meter data with a service provider. We assume that the top-k most frequently used appliances are specified by a service provider.

\subsection{Mapping the Individual Appliances' Energy Consumption into Range Levels}
\label{subsec:map}

Smart meters generally measure energy consumption within a certain range [0, max] at a specific time $t$. The `max' value is selected based on prior knowledge of energy consumption patterns, independent of the data.
It is reasonable to assume that energy data submitted by honest users will fall within the range.
In this step, we map energy consumption values (numeric) into distinct energy levels $L$ by dividing the maximum range of energy consumption [0, max] into smaller, distinct ranges. As appliances have varying energy consumption levels, different ranges were used to map the data. Specifically, we create a set of distinct ranges for energy levels $L$, $r$ = \{(0,$r_1$):$l_1$, ($r_1$, $r_2$):$l_2$,$\dots$, ($r_d$, max):$l_d$\} where the number of energy levels $L$ is equal to $d$, we label each range [$r_i$, $r_j$] as a distinct energy level $L$ = \{$l_1$, $l_2$, $\dots$, $l_d$\}. 
The service provider determines the number of levels ($d$) and the ranges ($r$). Still, all users must use the same configuration for energy consumption values to ensure uniform statistical analysis. 

For a user $u_i$, the energy consumption ($v_i$) for an appliance $a_i$ is mapped to the corresponding range ($r_i, r_j$), and the corresponding energy level, $l_i$, is generated. The aforementioned process is applied iteratively to all appliances that belong to a user (requiring users to report the energy consumption of all appliances listed by the service provider even if they don't have the appliance), which produces a set of energy levels for the user's energy consumption: \{$a_1$:$l_1$, $a_2$:$l_2$,$\dots$, $a_n$:$l_i$\}, where $n$ represents the total number of appliances. These energy levels can then be encoded to generate a flattened vector representing the energy consumption ranges of all appliances belonging to a user.

\subsection{Encoding, Vector Generation and Randomization}
%Energy Consumption Levels, Generating Flattened 1-D Vector, and Randomizing to Enforce DP}

After mapping the appliances' energy consumption values to energy levels, the discretized energy levels are encoded into binary vectors using binary encoding. This step can also reduce computation and communication costs in data transactions. The encoded energy levels for all appliances are then combined and flattened into a single 1-D vector, which is randomized to ensure privacy in the resulting binary vector.

\subsubsection{Binary Encoding}
\label{subsec:binaryenc}

Each appliance's energy consumption value is converted into a binary value with $d$ number of bits, where $d$ is the total number of energy levels. Each bit of the binary vector represents a specific energy level $l_i$.
If an appliance's energy consumption value falls within the range of a certain energy level $l_i$ (determined in step 1), a binary value will have a `1' in the corresponding bit position of $l_i$, and all other positions will have `0'.
For example, suppose an appliance's energy consumption belongs to $l_1$. In that case, the corresponding encoded value will be `$[0, 0, 0, 0, 0, 0, 0, 0, 0, 1]$' where the first-bit position is `1' (if the energy consumption value falls in $l_2$, the second-bit position will be `1') while the rest are `0's, given that $d$ is 10.

\subsubsection{1-D Vector Generation}

To prepare for the randomization process, all the binary values of the encoded appliance's energy consumption are combined into a single, flattened, long binary 1-D vector ($B$). 
$B$ has a length of $m$ = $d$ $\times$ $n$, where $d$ is the number of energy levels and $n$ is the number of appliances.

This step is necessary to ensure that when we distribute the privacy budget among all appliances while the utility is not adversely affected. In accordance with the composition property of DP (refer to Section \ref{subsec:composition}), when handling each appliance individually, a small fraction of the privacy budget will be allocated for randomizing the binary value of each appliance, potentially reducing utility. However, by combining all the encoded values of appliances into a single extended binary 1-D vector, the total privacy budget is fully utilized for all appliances. This approach not only enhances utility but also improves communication efficiency.

\vspace{-2mm}

\subsubsection{Randomization}

In the randomization step, we select Optimized Unary Encoding (OUE) over Unary Encoding (UE) (see Section~\ref{subsec: OUE}) to randomize $B$ to maintain a high utility of the randomized data, as UE can introduce undesirable levels of randomization due to the high sensitivity of the inputs when there are many appliances. 
The OUE approach randomizes the encoded bit values by flipping them from 0's to 1's or vice versa with a certain probability. 
This step results in a binary vector that may contain multiple 1s and 0s due to the randomization process, adding a layer of privacy protection to the encoded values. 
The probability of randomization, denoted as $p$, can be determined using Eq. (\ref{eq:randOUE}), where the probability of perturbing 0 to 1
 ($p_{0\to 1}$) is reduced by utilizing OUE (compared to UE, refer to Proof \ref{proof:UE}). The sensitivity of the encoded value for an individual appliance is 2, as the maximum possible difference of two binary inputs ($B_1$ and $B_2$, refer to Section \ref{subsec:ldp}) is 2 bits. Hence, when all appliance binary strings are combined into $B$, the sensitivity becomes $2 \times n$.

 \vspace{-3mm}

\begin{equation}
    \label{eq:randOUE}
    \begin{split}
     p =  \frac{1}{2}, \quad
     q =  \frac{1}{(1 + e^{\frac{\varepsilon}{2n/2}})}
     \end{split}
\end{equation}

Theorem \ref{theorm:LDPSmart} shows that {\mysys} satisfies differential privacy. The length of $B$ is $m = d \times n$, where $d$ represents the number of energy levels and $n$ represents the number of appliances. 

% \vspace{-3mm}

\begin{theorem}
\label{theorm:LDPSmart}

Let $v_1$ and $v_2$ be any bit positions of any two binary vectors $b_1$ and $b_2$, respectively, and let $B$ be a d-bit binary string output. When the probability of $B[v_1]$ being 1 given $v_1$ ($[\Pr[B[v_{1}]=1|v_{1}]$) is $\frac{1}{2}$, and the probability of $B[v_2]$ being 0 given $v_1$ ($\Pr[B[v_{2}]=0|v_{1}]$) is $\frac{ e^{\frac{\varepsilon}{2n/2}}}{1 +  e^{\frac{\varepsilon}{2n/2}}}]$, then the randomization model provides $\varepsilon$-LDP.

\begin{proof}
%Let $\varepsilon$ represent the privacy budget and the sensitivity of our approach is $2n$

\begin{align*}
\frac{\Pr[B|v_{1}]}{\Pr[B|v_{2}]} &= \frac{\prod_{i\varepsilon[d]}^{}\Pr[B[i]|v_{1}]}{\prod_{i\varepsilon[d]}^{}\Pr[B[i]|v_{2}]} \\
&\le \left[  \frac{\Pr[B[v_{1}] = 1|v_{1}]\Pr[B[v_{2}] = 0|v_{1}]}{\Pr[B[v_{1}] = 1|v_{2}]\Pr[B[v_{2}] = 0|v_{2}]}\right]^{2n/2} \\
&=\left( \frac{1/2}{\frac{1}{1+ e^{\frac{\varepsilon}{2n/2}}}}.\frac{\left( \frac{e^{\frac{\varepsilon}{2n/2}}}{1 + e^{\frac{\varepsilon}{2n/2}}}{} \right)}{1/2} \right)^{2n/2}  \\
&=e^{\varepsilon}
\end{align*}

\end{proof}
\end{theorem}

\subsection{Adaptive Budget Distribution}
%Protecting Privacy of Data Streams Using }

Since energy consumption data are collected over time, there is a potential risk of gradual privacy leakage over time. So, as shown in Figure \ref{fig:workflow}a, in step 3 of the workflow, the adaptive budget division methods LBU, LSP, LBD, and LBA are used to calculate the privacy budget for each timestamp within a window to enforce DP on smart meter stream data over time (refer to Section \ref{subsec:adaptive}).

Typically, LBD/LBA can be broken down into three components: private dissimilarity calculation, private strategy determination, and privacy budget allocation, as outlined in Figure \ref{fig:workflowadaptive} (refer to Appendix \ref{app:adaptive_lbd_lba}.

%As described in Section \ref{subsec:adaptive}, 
LBD/LBA dynamically selects between publication and approximation based on the dissimilarity $dis$ and potential publication error $err$ at each timestamp. If $dis < err$, the approximation is chosen, the current budget is set to 0, and the previous release is published directly. Otherwise, the perturbation strategy is used with the allocated budget ($\varepsilon_{t,2}$) for the randomization step (step 2 in Figure \ref{fig:workflow}).

Theorem \ref{thoe:lbd} shows that for stream data, the adaptation of LBD and LBA methods in {\mysys} satisfies $w$-event local differential privacy.

\begin{theorem}
    \label{thoe:lbd}
    \begin{proof}
        See Appendix Section \ref{subsec:appLBD}: Proof \ref{proof:lbd}
    \end{proof}
\end{theorem}

\subsection{Data Analysis Scenario by Service Provider} 

We use ``energy consumption aggregation and estimation of the Top-k commonly used appliances" as a data analysis scenario performed by the service provider.

As per the agreed-upon time interval for data collection, a user sends a randomized binary vector to the service provider. Upon receiving perturbed data from all users, the service provider aggregates the data and estimates the energy values of the top k appliances. {\mysys} employs LDP for each user, and all data instances are assumed to be independent (i.e., different users), resulting in a final privacy budget consumption that is the maximum of all privacy budgets used by each user at a time. 
%
%(max{$\varepsilon_1, \varepsilon_2, ..., \varepsilon_n$} - DP). 
%
A perturbed $B$ string is separated into individual randomized binary values for each appliance. 
%
%Since each $B'_i$ is within $L$ ($d$ number of energy levels), 
%
Consequently, the service provider can determine the energy consumption level $l_i$ for the top-k appliances by counting the frequency of each level from all users for that appliance.

\subsection{The Algorithm}

The exact steps for conducting LDP on disaggregated energy consumption data from individual household appliances using {\mysys} are outlined in Algorithm \ref{alg1}. Algorithm \ref{alg1} follows the detailed methodology discussed in Section \ref{sec:approach}. The later part of Table \ref{tab:TableOfNotations1} provides the notations and their corresponding descriptions used in Algorithm \ref{alg1}.

Algorithm \ref{alg3} outlines the process of server-side aggregation and estimation of energy levels for the top-k appliances from $k$ users. After data aggregation, an appliance may have multiple energy levels with a given count of users, as this can occur due to perturbation or differences in users' energy consumption patterns.
%(\{\{<$a_1$,$l_i$,$c_i$>, $\dots$, <$a_1$, $l_i$, $c_j$>\}, $\dots$, \{<$a_n$,$l_i$,$c_i$>, $\dots$, <$a_n$, $l_i$, $c_j$>\}\}).
% \begin{equation}

\begin{table}[ht]
\caption{Algorithm \ref{alg1} notations. Table \ref{tab:TableOfNotations} in the Appendix presents a more detailed version of notations.}
\centering 
\resizebox{0.5\textwidth}{!}{
\begin{tabular}{p{3.4cm}|p{5.2cm}}
\hline
\textbf{Notations}  & \textbf{Description} \\ \hline
% \multicolumn{2}{p{0.9\linewidth}}{ \centering\textbf{Notations for Algorithm \ref{alg1}}} \\ \hline
$n$ & Number of appliances \\ \hline
\{<$a_1$,$v_1$>, $\dots$, <$a_n$, $v_n$>\} &  $n$ Number of appliances and their corresponding energy values \\ \hline
 \{(0,$r_1$):$l_1$, $\dots$, ($r_d$, max):$l_d$\} & Energy ranges \\ \hline
 $d$ & Number of bits in the binary array, which represents the total number of energy levels \\ \hline
\end{tabular}
}
\label{tab:TableOfNotations1}
\end{table}

\section{Results and Discussion}
\label{sec:reanddisc}

In this section, we comprehensively evaluate the performance of {\mysys}%\footnote{The source code for {\mysys} will be released through a GitHub repository upon the acceptance of this work.} 
using the IDEAL Household Energy Dataset \cite{goddard2020ideal} and synthetic datasets to test {\mysys} against various data distributions. We employ multiple measurements (e.g., different window sizes, epsilon values, and levels) to support this evaluation.

The experiments were conducted on a Corsair Carbide Air 240 computer with a 4.20 GHz Intel(R) 8-Core i7-7700K processor and 32 GB of 2133 MT/s DDR4 memory. {\mysys} was implemented in Python, version 3.7. 

\subsection{The datasets}

The IDEAL dataset comprises detailed energy consumption data from 39 households in the UK, including usage information for 15 individual electrical appliances. We identified this dataset as the most extensive dataset available, encompassing a wide range of appliances and a reasonable number of users compared to other datasets. However, as this dataset is small to conduct the LDP approach, the dataset was augmented by synthesizing additional data points to emulate a sample size of 1000 users, utilizing a pattern consistent with the original IDEAL dataset. To ensure a consistent pattern during data augmentation, we adopt an approach that considers the appliance-wise mean, standard deviation, and lower-upper bounds of energy consumption for each user \cite{maharana2022review}. By focusing on appliance-wise patterns for each user, we can capture the individual usage patterns of each user more accurately. This approach allows us to preserve users' unique energy consumption patterns while augmenting the data. In the study conducted by Stadler et al.~\cite{stadler2022synthetic}, it was demonstrated that data augmentation techniques can maintain the authenticity of data. Hence, we assume that our augmentation methods are capable of preserving the quality of real data.

Next, the synthetic dataset was generated with varying distributions within the energy value range of [0, 3000] Watts. We chose this range based on our observations of the IDEAL data distribution. The generated data included normal, uniform, and skewed (left and right) distributions to simulate varying energy consumption patterns with 10,000 simulated users.

\subsection{Experiments for Utility Analysis}
\label{app:utility}

We quantitatively assessed the effectiveness of {\mysys} by comparing the aggregated perturbed data from all users to their corresponding true appliance energy levels. % using the Kruskal--Wallis test \cite{mckight2010kruskal}. 
For that, we chose the Kruskal-Wallis test \cite{mckight2010kruskal} because it is specifically designed for non-normal distributions (which is true for a majority of real-world datasets including energy data) \cite{mckight2010kruskal}. This test is non-parametric, meaning that it does not assume that the data follow a specific distribution, and it is particularly useful when the data violate the assumptions of normality and equal variance. 
Therefore, as follows, we applied the Kruskal--Wallis test to assess the similarity of the underlying distributions of the true and perturbed histograms.
% (refer to Appendix \ref{app:utility}, for further details on this evaluation). 

Consider a scenario where $k$ users sharing perturbed energy consumption data for $n$ appliances at a specific time point $t$,  

\begin{itemize}
     \item $n$ number of appliances and their energy consumption values = \{<$a_1$,$v_1$>, \dots, <$a_n$, $v_n$>\}
    \item $d$ levels of energy and range of the level = \{(0,$r_1$):$l_1$, ($r_1$, $r_2$):$l_2$,\dots, ($r_d$, max):$l_d$\} (max- maximum energy value) 
    \raggedright
    \item For a user $u_{1}$, mapped <appliance, energyLevel> set ($c_t$) = $\left\{ <a_{1}, l_{1}>, \dots, <a_{n}, l_{i}> \right\}$
\end{itemize}

\begin{figure}[h]
     \centering
     \begin{subfigure}[b]{0.20\textwidth}
         \centering
         \includegraphics[width=\textwidth]{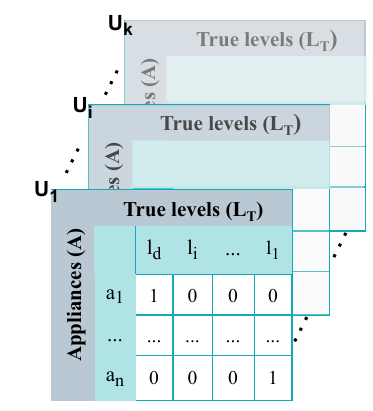}
         \caption{True data}
        \label{fig:true}
     \end{subfigure}
     \hfill
     \begin{subfigure}[b]{0.20\textwidth}
         \centering
         \includegraphics[width=\textwidth]{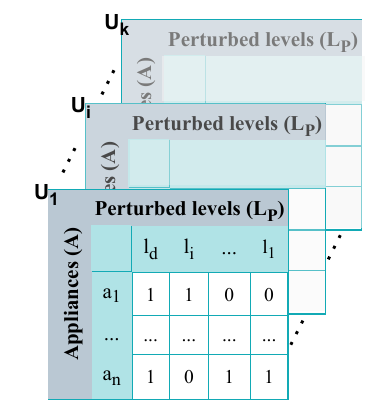}
         \caption{Perturbed data}
         \label{fig:per}
     \end{subfigure}
     \caption{Comparison of true and perturbed energy consumption data for utility assessment for all users.}
 \end{figure}

The utility was assessed for all users' aggregated true data for each appliance (refer to Figure \ref{fig:true}) with perturbed data (refer to Figure \ref{fig:per}) using the Kruskal-Wallis test. 
% The Kruskal-Wallis test is used to compare different distributions.
The Kruskal-Wallis test ranks the observations from both distributions (true and perturbed) and determines if the samples come from the same distribution.% by calculating a test statistic.
A p-value greater than the established threshold (0.05) for statistical significance indicates insufficient evidence to reject the null hypothesis of no difference between the two distributions. That implies that the two distributions (true and perturbed) show similarity.

\subsubsection{Experiments with IDEAL dataset}

For the experiments, we utilize a default configuration of 1000 simulated users, 15 distinct appliances, and 10 quantization levels. % within the range of [0-2500].
The hyperparameters are selected independently from data, which are generally defined by service providers based on prior knowledge. In real-life scenarios, service providers often select the energy level range based on their previous experiences and knowledge gained from analyzing prior data. Our experiments encompass a variety of level ranges independent of the data dynamics. 

Data are distributed among the simulated users and shared once a day within a sliding window of 3 days. We chose to focus on 15 appliances in our experiments due to the common range of household possession, which typically falls between 10 and 17 appliances \cite{won2014survey}. Literature studies \cite{statistaHouseholdAppliances, AGL} highlight that the vast majority of energy usage comes from roughly 10-12 appliance categories, which account for nearly 99\% of total energy consumption \cite{renewableenergyworldHomeAppliances}. Our approach can maintain utility for a larger number of appliances, however, it requires a large number of users. %We performed quantization of the energy level range for each bin.
% based on the distribution of our dataset. 

Firstly, we evaluated the performance of {\mysys} using various adaptive budget division methods such as LBU, LSP, LBD, and LBA with a privacy budget of 10 (a higher budget was chosen as we needed to maintain the utility of 15 appliances in a window and our dataset is relatively small in size.). In our scenario, allocating a maximum of $\varepsilon = 10/15 = 0.
66$ for each appliance, on average, is reasonable. This allocation aligns with the approach used by Apple as outlined in their differential privacy implementation \cite{apple}, where they utilize a privacy budget with $\varepsilon$ ranging from 4 (Lookup Hints) to 8 (Safari Auto-play intent detection) for their products with millions of users, collecting data once or twice per day.
Considering our limited dataset of 1000 users, 15 appliances, and a window size of 3, our proposed privacy budget of $\varepsilon = 10$ is reasonable.

Figure \ref{fig:real:methodComparison} displays the averaged p-values obtained from the Kruskal--Wallis test for the LBU, LSP, LBD, and LBA methods. The p-values were calculated for 15 appliances over 10 iterations. We computed p-values for each appliance by comparing the true and perturbed data and then calculated the average p-value for all appliances. The results were visualized in a graph (refer to Figure \ref{fig:real:methodComparison}). In most cases, the p-values are less than 0.05, indicating less similarity. This is expected because our dataset is small, has a limited number of users, and is completely randomized. DP fundamentally requires a large amount of data from many users for a better approximation. However, as shown in Figure \ref{fig:real:levelComparison}, when the number of levels is 5, most appliances show p-values greater than 0.05 (more results are available in the Appendix, Table \ref{tab:data_report}), indicating a similar pattern between the perturbed and true data. 
Furthermore, the application of top-k retrieval on perturbed data (discussed in Section \ref{subse:serversideanalysis}) shows promising performance with an average hit rate of 5.7, validating the performance capabilities of our method.

The LSP method outperforms other methods in the given dataset. This can be attributed to the characteristics of the dataset, which contains numerous instances of zero values due to the absence of energy consumption for many devices over time. The LSP method effectively utilizes a higher budget for each publication by sampling other releases in this dataset because the statistics show minimal variation across neighboring timestamps in the IDEAL dataset. However, it should be noted that the LBD and LBA methods outperform LSP in our synthetic dataset (refer to Figure \ref{fig:distributions}) due to their data-dependent sampling, dynamic budget allocation, and approximation strategy. The LBA method performs well in the IDEAL dataset, with only a slight difference compared to LSP. This is because LBA effectively allocates the budget over multiple publications. It achieves this through an initial uniform budget distribution and subsequent dynamic budget absorption. The performance maintenance of LBA becomes more evident when we increase the window size, as shown in Figure \ref{fig:real:windowComparison}. Moreover, the LSP method shows minimal variation across different configurations, indicating that it tends to sample timestamps without considering the underlying data distribution. This approach may not capture users' diverse energy consumption patterns.
Hence, for subsequent experiments, LBA was chosen as our default method with different configurations to assess the influence of other parameters in the utility.

We conducted the utility analysis utilizing the same configuration, except for the window size $w$, by incrementing the window size from 2 to 7.
Figure \ref{fig:real:windowComparison} shows that a larger window size gives lower p-values, indicating more dissimilarity. The results are consistent with our theoretical utility analysis, as the privacy budget ($\varepsilon$) is divided by $w$ in LBD and LBA approaches. Besides, Figure \ref{fig:real:epsilonComparison} is consistent with the DP process, where high $\varepsilon$ values tend to decrease dissimilarity. We then examined the relationship between the number of energy levels (which also changes the ranges) and utility. Figure \ref{fig:real:levelComparison} shows that when the number of levels is 5, for most of the appliances, we achieve a p-value greater than 0.05, indicating a similar pattern between the perturbed and true data. Increasing the number of levels significantly reduces the utility of our approach, as the probability of falling into different levels increases, but the privacy budget ($\varepsilon$ = 10) remains unchanged. This means there is a certain probability of flipping the energy levels from 0 to 1 or vice versa. Further, as the number of levels increases, there is more randomization over many levels, leading to a decrease in utility.

\pgfplotsset{compat=newest}
\begin{figure}[!t]
  \centering
  \begin{subfigure}[b]{0.23\textwidth}   
  \begin{tikzpicture}
    \begin{axis}[
        % /pgf/number format/1000 sep={},
        width=0.8\textwidth,
        height=1.2in,
        scale only axis,
        clip=false,
        separate axis lines,
        axis on top,
        xmin=0,
        xmax=5,
        xtick={1,2,3,4},
        x tick style={draw=none},
        yticklabel style={font=\scriptsize},
        xticklabel style={font=\scriptsize}, % Adjust the font size here
        xticklabels={LBU, LSP, LBD, LBA},
        ymin=0,
        y label style={at={(-0.20,0.25)}, anchor=west},
        ylabel style={font=\small},
        ylabel={p-values},
        xlabel style={font=\small},
        xlabel={Adaptive methods},
        every axis plot/.append style={
          ybar,
          bar width=.5,
          bar shift=0pt,
          fill
        }
      ]
      \addplot[colorlbu]coordinates {(1,0.00044)};
      \addplot[colorlsp]coordinates{(2,0.00111)};
      \addplot[colorlbd]coordinates{(3,0.00058)};
      \addplot[colorlba]coordinates{(4,0.00082)};
    \end{axis}
  \end{tikzpicture}
  \caption{Different adaptive budget methods} 
  \label{fig:real:methodComparison}
  \end{subfigure}
  \begin{subfigure}[b]{0.23\textwidth}
  \begin{tikzpicture}
    \begin{axis}[
        % /pgf/number format/1000 sep={},
        width=0.8\textwidth,
        height=1.2in,
        scale only axis,
        clip=false,
        separate axis lines,
        axis on top,
        xmin=0,
        xmax=5,
        xtick={1,2,3,4},
        x tick style={draw=none},
        yticklabel style={font=\scriptsize},
        xticklabel style={font=\scriptsize},
        xticklabels={w=2, w=3, w=5, w=7},
        ymin=0,
        y label style={at={(-0.18,0.25)}, anchor=west},
        ylabel style={font=\small},
        ylabel={p-values},
        xlabel style={font=\small},
        xlabel={Window sizes},
        every axis plot/.append style={
          ybar,
          bar width=.5,
          bar shift=0pt,
          fill
        }
      ]
      \addplot[colorw2]coordinates {(1,0.00097)};
      \addplot[colorw3]coordinates{(2,0.00082)};
      \addplot[colorw5]coordinates{(3,0.00066)};
      \addplot[colorw7]coordinates{(4,0.00059)};
    \end{axis}
  \end{tikzpicture}
  \caption{Different number of window sizes}
  \label{fig:real:windowComparison}
  \end{subfigure}
  \begin{subfigure}[b]{0.23\textwidth}
  \begin{tikzpicture}
    \begin{axis}[
        % /pgf/number format/1000 sep={},
        width=0.8\textwidth,
        height=1.2in,
        scale only axis,
        clip=false,
        separate axis lines,
        axis on top,
        xmin=0,
        xmax=5,
        xtick={1,2,3,4},
        x tick style={draw=none},
        yticklabel style={font=\scriptsize},
        xticklabel style={font=\scriptsize}, % Adjust the font size here
        xticklabels={$\varepsilon$=5, $\varepsilon$=10, $\varepsilon$=15, $\varepsilon$=20},
        ymin=0,
        y label style={at={(-0.20,0.25)}, anchor=west},
        ylabel style={font=\small},
        ylabel={p-values},
        xlabel style={font=\small},
        xlabel={Privacy budgets},
        every axis plot/.append style={
          ybar,
          bar width=.5,
          bar shift=0pt,
          fill
        }
      ]
      \addplot[color16]coordinates {(1,0.00061)};
      \addplot[color17]coordinates{(2,0.00082)};
      \addplot[color18]coordinates{(3,0.00097)};
      \addplot[color19]coordinates{(4,0.00106)};
    \end{axis}
  \end{tikzpicture}
  \caption{Different privacy budgets} 
  \label{fig:real:epsilonComparison}
  \end{subfigure}
  \begin{subfigure}[b]{0.23\textwidth}
  \begin{tikzpicture}
    \begin{axis}[
        % /pgf/number format/1000 sep={},
        width=0.82\textwidth,
        height=1.2in,
        scale only axis,
        clip=false,
        separate axis lines,
        axis on top,
        xmin=0,
        xmax=5,
        xtick={1,2,3,4},
        x tick style={draw=none},
        yticklabel style={font=\scriptsize},
        xticklabel style={font=\scriptsize},
        xticklabels={l=5, l=7, l=10, l=15},
        ymin=0,
        y label style={at={(-0.12,0.25)}, anchor=west},
        ylabel style={font=\small},
        ylabel={p-values},
        xlabel style={font=\small},
        xlabel={Levels},
        every axis plot/.append style={
          ybar,
          bar width=.5,
          bar shift=0pt,
          fill
        }
      ]
      \addplot[color20]coordinates {(1,0.04713)};
      \addplot[color21]coordinates{(2,0.01054)};
      \addplot[color22]coordinates{(3,0.00082)};
      \addplot[color23]coordinates{(4,0.00082)};
    \end{axis}
  \end{tikzpicture}
  \caption{Different number of levels} 
  \label{fig:real:levelComparison}
  \end{subfigure}
 \caption{The utility analysis over different configurations: the average p-values for all appliances of Kruskal--Wallis similarity test in the IDEAL dataset. Except for the variations in each specific scenario represented in the graphs, the default configuration is: w = 3, adaptive method = LBA, number of appliance = 12, $\varepsilon$ = 10, l = 10}
 \label{fig:real_graphs}
\end{figure}

\paragraph{\textbf{Service Provider-Side Application of Perturbed Data}}
\label{subse:serversideanalysis}

In addition to conducting the Kruskal-Wallis similarity tests, we evaluated our approach by applying various service-provider-side post-processing applications, such as top-k appliance inference, and analyzing appliance usage patterns on the perturbed data %to evaluate the usefulness of our approach in relation to the true data distribution.  

Our first application involves calculating the top-k appliances within an aggregated dataset comprising data from all users. We computed the total energy consumption for each appliance over the experimented days (30 days in our experiments) and sorted the list in descending order to retrieve top-k appliances. Figures \ref{fig:topk:true} and \ref{fig:topk:perturbed10} showcase the true and perturbed data results. 
Despite having limited data and operating in a constrained environment with 15 appliances, a window size of 3, and a privacy parameter of 10, our approach showed promising results in predicting the top 10 appliances within 15 appliances. In our evaluation (refer to Figures \ref{fig:topk:true} and \ref{fig:topk:perturbed10}), our approach demonstrated successful identification of the 8 appliances within the top 10 appliances and displayed minimal rank changes beyond that point.
Using the hit rate definition ($hit count / total count$), our approach achieved a hit rate of 8/10, indicating that it accurately predicted 8 out of the top 10 appliances. The `hit count' denotes the number of times the top-k appliances are accurately identified with the correct rank. 
To obtain a more consistent average hit rate, we reran our algorithm 100 times and observed an average hit rate of 5.7 (median: 6), with varying hit rates ranging from 4 to 8. It is noteworthy that the top appliances consistently yielded accurate results.

\begin{figure}[!t]
\centering
  \begin{subfigure}[b]{0.43\textwidth}
  \begin{tikzpicture}[scale=0.9]
  \tikzstyle{every node}=[font=\scriptsize]
  \tikzset{
     lines/.style={draw=none},
    }
        \pie[color={color1, color5, color2, color3, color4, color6, color7, color8, color9, color10, color11, color12, color13, color14, color15}, style={lines}]{27.1/A7, 18.1/A6, 13.4/A2, 9.1/A3, 8.4/A4, 6.5/A1, 4.1/A11, 3.1/A9, 2.9/A8, 2.7/A13, 1.1/A5, 1.1/A15, 0.8/A10, 0.8/A14, 0.7/A12}
        % \pie[color={color1, color5, color2, color3, color4, color6, color7, color8, color9, color10, color11}, style =very thin]{27.1/A7, 18.1/A6, 13.4/A2, 9.1/A3, 8.4/A4, 6.5/A1, 4.1/A11, 3.1/A9, 2.9/A8, 2.7/A13, 4.5/Other}
    \end{tikzpicture}
  \caption{True data}
  \label{fig:topk:ex2true}
  \end{subfigure}
  \begin{subfigure}[b]{0.43\textwidth}
 \begin{tikzpicture}[scale=0.9]

  \tikzstyle{every node}=[font=\scriptsize]
   \tikzset{
     lines/.style={draw=none},
    }
        \pie[color={color1, color5, color2, color3, color4, color6, color7, color8, color9, color10, color11, color12, color13, color14, color15}, style={lines}]{9.8/A7, 8.7/A6, 7.4/A2, 7.1/A3, 6.7/A4, 6.6/A1, 6.4/A11, 6.3/A9, 6.1/A14, 6.0/A8, 5.8/A5, 5.8/A15, 5.8/A10, 5.7/A13, 5.7/A12}
        % \pie[color={color1, color5, color2, color3, color4, color6, color7, color8, color9, color10}, style =very thin]{9.8/A7, 8.7/A6, 7.4/A2, 7.1/A3, 6.7/A4, 6.6/A1, 6.4/A11, 6.3/A9, 6.1/A14, 6.0/A8, 28.8/Other}
    \end{tikzpicture}
    \caption{Perturbed data}
     \label{fig:topk:ex2perturbed10}
  \end{subfigure}
 \caption{Service provider-side application on IDEAL dataset: Impact of individual appliances on overall energy consumption}
        \label{fig:impact:addtional_experiments}
\end{figure}
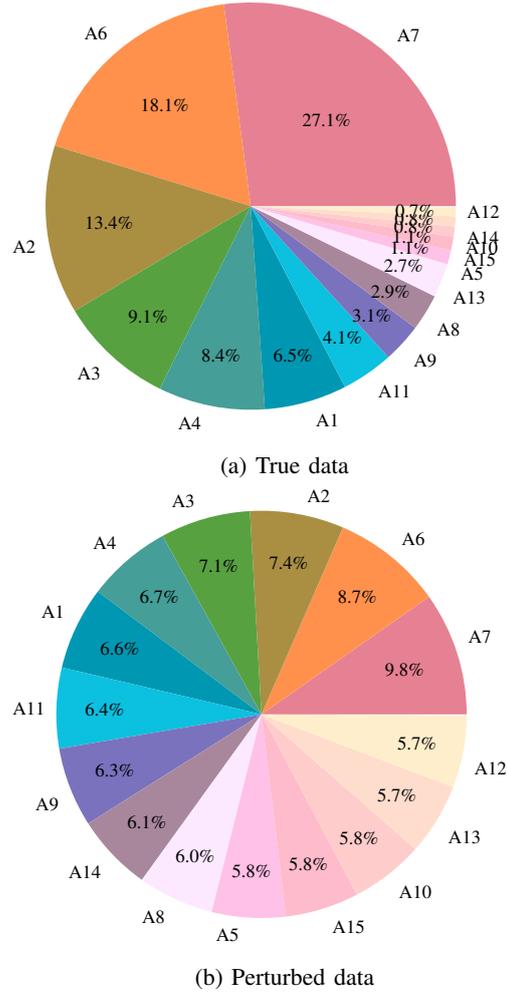

Next, we examined the impact of individual appliances on overall energy consumption during the given time period, as shown in Figures \ref{fig:topk:ex2true} and \ref{fig:topk:ex2perturbed10}. Although true and perturbed data percentages differ in both scenarios, we observe a consistent pattern. For instance, in the pie chart, appliances `A7', `A6', and `A2' have a greater impact on total energy consumption, while appliances `A5' and `A10' have a lesser impact. This pattern remains evident even with perturbed data.

\begin{figure}[!t]
  % \centering
  \begin{subfigure}[b]{0.23\textwidth}   
  \begin{tikzpicture}
    \begin{axis}[
        % /pgf/number format/1000 sep={},
        width=0.8\textwidth,
        height=1.2in,
        at={(0.758in,0.981in)},
        scale only axis,
        clip=false,
        separate axis lines,
        axis on top,
        xmin=0,
        xmax=11,
        xtick={1,2,3,4,5,6,7,8,9,10,11},
        x tick style={draw=none},
        y tick style={draw=none},
        yticklabel style={font=\scriptsize},
        xticklabel style={font=\scriptsize}, % Adjust the font size here
        xticklabels={A7, A6, A2, A3, A4, A1, A11, A9, A8, A13},
        x tick label style={rotate=90,anchor=east},
        ymin=0,
        y label style={at={(-0.01,0.5)}},
        ylabel style={font=\small},
        ylabel={Avg. energy consumption},
        xlabel style={font=\small},
        xlabel={Appliances},
         yticklabels={}, % Add this line to hide y-tick values
        every axis plot/.append style={
          ybar,
          bar width=.8,
          bar shift=0pt,
          fill
        }
      ]
      \addplot[color1]coordinates {(1,27.1)};
      \addplot[color5]coordinates{(2,18.1)};
      \addplot[color2]coordinates{(3,13.4)};
      \addplot[color3]coordinates{(4,9.1)};
      \addplot[color4]coordinates {(5,8.4)};
      \addplot[color6]coordinates{(6,6.5)};
      \addplot[color7]coordinates{(7,4.1)};
      \addplot[color8]coordinates{(8,3.1)};
      \addplot[color9]coordinates{(9,2.9)};
      \addplot[color10]coordinates{(10,2.7)};
    \end{axis}
  \end{tikzpicture}
     \caption{True data}
     \label{fig:topk:true}
  \end{subfigure}
  \begin{subfigure}[b]{0.23\textwidth}   
  \begin{tikzpicture}
    \begin{axis}[
        % /pgf/number format/1000 sep={},
        width=0.8\textwidth,
        height=1.2in,
        at={(0.758in,0.981in)},
        scale only axis,
        clip=false,
        separate axis lines,
        axis on top,
        xmin=0,
        xmax=11,
        xtick={1,2,3,4,5,6,7,8,9,10,11},
        x tick style={draw=none},
        y tick style={draw=none},
        yticklabel style={font=\scriptsize},
        xticklabel style={font=\scriptsize}, % Adjust the font size here
        xticklabels={A7, A6, A2, A3, A4, A1, A11, A9, A14, A8},
         x tick label style={rotate=90,anchor=east},
        ymin=0,
        y label style={at={(-0.01,0.5)}},
        ylabel style={font=\small},
        ylabel={Avg. energy consumption},
        xlabel style={font=\small},
        xlabel={Appliances},
         yticklabels={}, % Add this line to hide y-tick values
        every axis plot/.append style={
          ybar,
          bar width=.8,
          bar shift=0pt,
          fill
        }
      ]
      \addplot[color1]coordinates {(1,9.8)};
      \addplot[color5]coordinates{(2,8.7)};
      \addplot[color2]coordinates{(3,7.4)};
      \addplot[color3]coordinates{(4,7.1)};
      \addplot[color4]coordinates {(5,6.7)};
      \addplot[color6]coordinates{(6,6.6)};
      \addplot[color7]coordinates{(7,6.4)};
      \addplot[color8]coordinates{(8,6.3)};
      \addplot[color9]coordinates{(9,6.1)};
      \addplot[color10]coordinates{(10,5.9)};
    \end{axis}
  \end{tikzpicture}
     \caption{Perturbed data}
     \label{fig:topk:perturbed10}
  \end{subfigure}
   \caption{Service provider-side application on IDEAL dataset: Top-k appliance distribution}
    \label{fig:topk:addtional_experiments}
\end{figure}
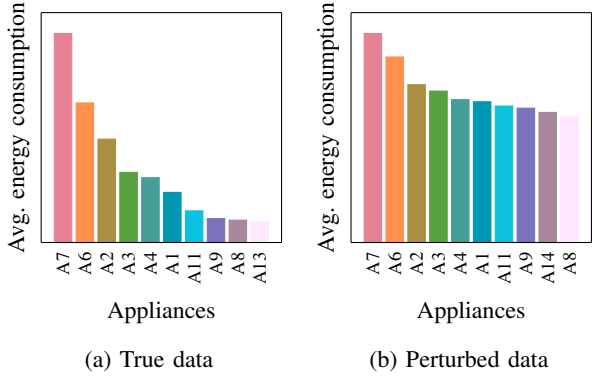

In summary, our experiment demonstrates that even with a limited dataset and privacy-constrained environment, our approach successfully preserves individual user privacy while holding utility by maintaining consistent patterns on an aggregated level. 

\subsubsection{Experiments with Synthetic Data}

In the experimental evaluation utilizing synthetic data of varying distributions, the default configuration was utilized, which consists of a total of 10,000 simulated users, 15 distinct appliances, 10 quantization levels, a privacy budget ($\varepsilon$) of 2 (we further restricted the environment), and a window size of 2.

We evaluated the utility of {\mysys} for different distributions using LBU, LSP, LBD, and LBA methods, which are shown in Figure \ref{fig:distributions}. On average, all methods performed similarly for most of the distributions; however, we observed a significant disparity in skewed distributions as utilizing LBD could potentially save the privacy budget when releases similar to past releases occur. The reason behind the better performance of the LBD method compared to LBA is the small window size ($w$) selected for our approach due to limited data. This allows LBD to maintain a reasonable privacy budget $\varepsilon$ as it decreases exponentially over a window. A smaller window size results in a slower exponential decrease in the privacy budget. %Hence, for subsequent experiments, LBD was chosen as our default method. 

The experiments reveal that LBD and LBA methods perform optimally across various distributions, while LBU and LSP excel in cases like uniform and normal distributions with minimal data fluctuations. However, since energy consumption data may exhibit fluctuations and not strictly adhere to specific distributions, employing LBD (for small window sizes) or LBA (for large window sizes) is preferable to balance utility and privacy preservation.

We further conducted additional experiments on synthetic data to investigate the impact of the number of users and appliances using the LBD method in Appendix \ref{sec:more_results}.

\pgfplotsset{compat=newest}
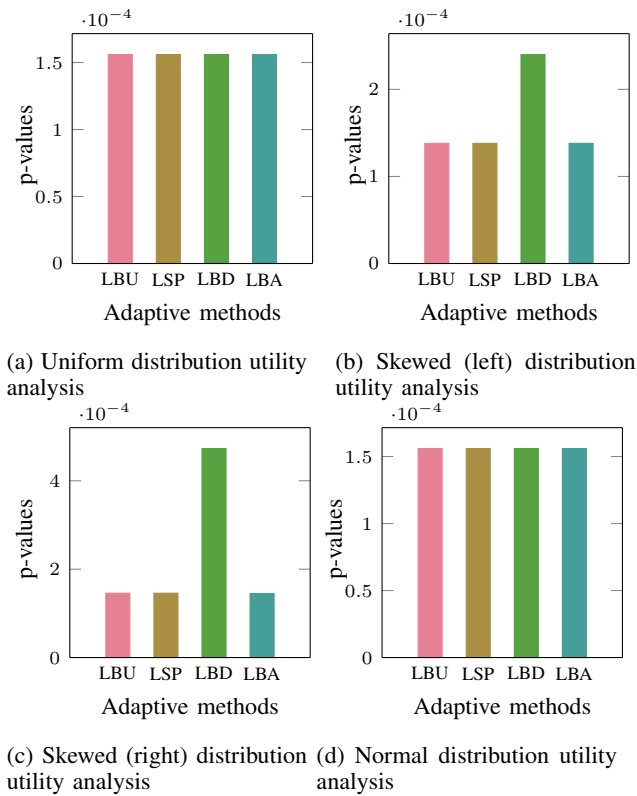
\begin{figure}[!t]
  % \centering
  \begin{subfigure}[b]{0.23\textwidth}   
  \begin{tikzpicture}
    \begin{axis}[
        % /pgf/number format/1000 sep={},
        width=0.8\textwidth,
        height=1.2in,
        scale only axis,
        clip=false,
        separate axis lines,
        axis on top,
        xmin=0,
        xmax=5,
        xtick={1,2,3,4},
        x tick style={draw=none},
        yticklabel style={font=\scriptsize},
        xticklabel style={font=\scriptsize}, % Adjust the font size here
        xticklabels={LBU, LSP, LBD, LBA},
        ymin=0,
        y label style={at={(-0.19,0.30)}, anchor=west},
        ylabel style={font=\small},
        ylabel={p-values},
        xlabel style={font=\small},
        xlabel={Adaptive methods},
        every axis plot/.append style={
          ybar,
          bar width=.5,
          bar shift=0pt,
          fill
        }
      ]
      \addplot[color1]coordinates {(1,0.000156)};
      \addplot[color2]coordinates{(2,0.000156)};
      \addplot[color3]coordinates{(3,0.000156)};
      \addplot[color4]coordinates{(4,0.000156)};
    \end{axis}
  \end{tikzpicture}
  \caption{Uniform distribution utility analysis} 
  \label{fig:uniform:methodComparison}
  \end{subfigure}
  \begin{subfigure}[b]{0.23\textwidth}
  \begin{tikzpicture}
    \begin{axis}[
        % /pgf/number format/1000 sep={},
        width=0.8\textwidth,
        height=1.2in,
        scale only axis,
        clip=false,
        separate axis lines,
        axis on top,
        xmin=0,
        xmax=5,
        xtick={1,2,3,4},
        x tick style={draw=none},
        yticklabel style={font=\scriptsize},
        xticklabel style={font=\scriptsize},
        xticklabels={LBU, LSP, LBD, LBA},
        ymin=0,
        y label style={at={(-0.14,0.30)}, anchor=west},
        ylabel style={font=\small},
        ylabel={p-values},
        xlabel style={font=\small},
        xlabel={Adaptive methods},
        every axis plot/.append style={
          ybar,
          bar width=.5,
          bar shift=0pt,
          fill
        }
      ]
      \addplot[color1]coordinates {(1,0.000138)};
      \addplot[color2]coordinates{(2,0.000138)};
      \addplot[color3]coordinates{(3,0.000240)};
      \addplot[color4]coordinates{(4,0.000138)};
    \end{axis}
  \end{tikzpicture}
  \caption{Skewed (left) distribution utility analysis}
  \label{fig:skew:methodComparison}
  \end{subfigure}
  \begin{subfigure}[b]{0.23\textwidth}
  \begin{tikzpicture}
    \begin{axis}[
        % /pgf/number format/1000 sep={},
        width=0.8\textwidth,
        height=1.2in,
        scale only axis,
        clip=false,
        separate axis lines,
        axis on top,
        xmin=0,
        xmax=5,
        xtick={1,2,3,4},
        x tick style={draw=none},
        yticklabel style={font=\scriptsize},
        xticklabel style={font=\scriptsize}, % Adjust the font size here
        xticklabels={LBU, LSP, LBD, LBA},
        ymin=0,
        y label style={at={(-0.18,0.30)}, anchor=west},
        ylabel style={font=\small},
        ylabel={p-values},
        xlabel style={font=\small},
        xlabel={Adaptive methods},
        every axis plot/.append style={
          ybar,
          bar width=.5,
          bar shift=0pt,
          fill
        }
      ]
      \addplot[color1]coordinates {(1,0.0001457)};
      \addplot[color2]coordinates{(2,0.0001456)};
      \addplot[color3]coordinates{(3,0.000473)};
      \addplot[color4]coordinates{(4,0.000145)};
    \end{axis}
  \end{tikzpicture}
  \caption{Skewed (right) distribution utility analysis} 
  \label{fig:skewp:methodComparison}
  \end{subfigure}
  \begin{subfigure}[b]{0.23\textwidth}
  \begin{tikzpicture}
    \begin{axis}[
        % /pgf/number format/1000 sep={},
        width=0.80\textwidth,
        height=1.2in,
        scale only axis,
        clip=false,
        separate axis lines,
        axis on top,
        xmin=0,
        xmax=5,
        xtick={1,2,3,4},
        x tick style={draw=none},
        yticklabel style={font=\scriptsize},
        xticklabel style={font=\scriptsize},
        xticklabels={LBU, LSP, LBD, LBA},
        ymin=0,
        y label style={at={(-0.19,0.30)}, anchor=west},
        ylabel style={font=\small},
        ylabel={p-values},
        xlabel style={font=\small},
        xlabel={Adaptive methods},
        every axis plot/.append style={
          ybar,
          bar width=.5,
          bar shift=0pt,
          fill
        }
      ]
      \addplot[color1]coordinates {(1,0.000156)};
      \addplot[color2]coordinates{(2,0.000156)};
      \addplot[color3]coordinates{(3,0.000156)};
      \addplot[color4]coordinates{(4,0.000156)};
    \end{axis}
  \end{tikzpicture}
  \caption{Normal distribution utility analysis} 
  \label{fig:normal:levelComparison}
  \end{subfigure}
 \caption{The Kruskal--Wallis similarity test p values for each appliance in uniform, normal, positive, and negative skewed distributions (in synthetic data)}
  \label{fig:distributions}
\end{figure}

\vspace{-0.5mm}

\subsection{Computational Complexity Analysis}
\label{subsec:computationalcomplex}

We assess the computational complexity of our approach in terms of the algorithm's time requirements. We utilize the appropriate notation ($O$) to describe the algorithm's computational complexity growth rate. 

At the users' end, smart meters simply perform real number energy consumption mapping and encoding to predefined range levels and budget calculations involving addition, subtraction, and multiplication in iterative loops for all appliances over time. 
Part I of the algorithm involves mapping the energy consumption of appliances to given range levels. According to Algorithm \ref{alg1}, the overall computational complexity is $O(n^3)$, representing the product of the complexities of the time dimension $O(t)$, the appliance dimension $O(n)$, and the range levels dimension $O(d)$. However, it is important to note that the algorithm operates on a single timestamp at a time, so we will now focus on analyzing the computational complexity for a single timestamp. This step has a computational complexity of $O(n) \times O(d)$, indicating a quadratic growth rate for a specific timestamp. The complexity is quadratic because it involves iterating over $n$ appliances and performing operations related to $d$ range levels.

Part II of the algorithm involves encoding and creating a flattened vector with a time complexity of $O(n)$. This indicates a linear growth rate, as the time taken for this step increases linearly with the number of appliances ($n$). Part III of the algorithm involves iterating over timestamps. However, the time taken for each iteration of the loop does not depend on the input size. Therefore, the computational complexity for this part is $O(1)$, indicating constant computational complexity. So, on the client side, the overall complexity is $O(n^2)$.

Service providers handle the aggregation part; the operations involve decoding and addition over multiple users. The computational complexity for this part can be represented as $O(k)$, where $k$ represents the number of users. 

\subsection{Benchmarking}

Existing methods primarily apply DP for energy consumption data aggregation. However, these methods fall short when extended to the individual appliance level because they may expose the appliance-level energy usage patterns over time in the context of data streams. Moreover, our approach is the first to integrate appliance-wise energy consumption, streaming data, and top-k appliance inference together. This unique combination renders direct comparisons with existing approaches (e.g., existing frequency estimation protocols) challenging. Hence, to ensure a pertinent and equitable assessment, we chose the existing well-known LDP algorithms, Laplace ~\cite{hassan2019differential, hossain2021cost}, Gamma, Gaussian, and Exponential mechanisms for benchmarking. 
% These previous studies primarily focus on total aggregation rather than appliance-wise information. 
It is worth noting that previous research predominantly emphasizes total aggregation rather than appliance-specific disaggregated data. To ensure a fair comparison, we implemented these approaches for disaggregated data.

In all mentioned benchmarking approaches, we apply a fixed privacy budget ($\varepsilon = 1$) at each time instant instead of considering an adaptive budget as in {\mysys}. The reason for this differentiation is that the existing LDP methods tailored for protecting smart meter privacy do not consider the streaming characteristics of energy data. Compared to our approach, the allocated privacy budget in the benchmarking methods is significantly higher, as it is allocated for each appliance. However, in the {\mysys} approach, we allocate a privacy budget of 10 for a 3-day window, considering a total of 15 appliances. Table \ref{tab:quantitative} quantitatively analyzes both the benchmarking and {\mysys} approaches using the IDEAL dataset in terms of time complexity, computational complexity, and hit rate. The recorded values were obtained by averaging 100 iterations.

\begin{table}[ht]
\caption{Quantitative Comparison of {\mysys} against the existing benchmarking approaches}
\label{tab:quantitative}
\centering
\begin{tabular}{p{1.6cm}|p{1.4cm}|p{2.5cm}|p{0.8cm}}
\hline

\textbf{Approach} & \textbf{Time complexity} & \textbf{Computational complexity} & \textbf{Average hit rate}\\ \hline
Laplace  & 0.06ms  & Client: $O(n)$, service provider: $O(n^2)$ & 1.98 \\ \hline
Gamma & 0.11ms & Client: $O(n)$, service provider: $O(n^2)$ & 2.09 \\ \hline
Gaussian  & 0.08ms & Client: $O(n)$, service provider: $O(n^2)$ & 2.075 \\ \hline
Exponential & 0.09ms & Client: $O(n)$, service provider: $O(n^2)$ & 2.1 \\ \hline
\textit{LDP-SmartEnergy} & 0.009ms & Client: $O(n^2)$, service provider: $O(k)$ & 5.7 \\ \hline
\multicolumn{4}{c}{*Number of concurrent jobs = number of appliances = 15} \\ \hline

\end{tabular}

\end{table}

In terms of time complexity, our initial experiments involved adding noise sequentially to all the appliances. However, since our {\mysys} approach perturbs all appliances simultaneously by combining them, we conducted concurrent implementations to ensure a fair comparison of time complexity for our benchmarking approaches. We set the number of concurrent jobs to match the number of appliances (n = 15) in this scenario. The time taken for concurrent noise addition by the benchmarking approaches is presented in Table \ref{tab:quantitative}. Notably, even in the concurrent noise addition scenarios, the time complexity of our {\mysys} approach is lower than that of other approaches. In the benchmarking mechanisms, managing data streams from appliances, per the composition theory, can result in a sophisticated structure that can become highly inefficient, as it handles each appliance independently. In contrast, our approach concurrently addresses the protection of all appliances. Due to this reason, our approach achieves a lower time complexity compared to other mechanisms. 

Regarding computational complexity, our algorithm shows a user-side complexity of $O(n^2$), while the service provider-side complexity is $O(k)$ (discussed in \ref{subsec:computationalcomplex}. The user-side's $O(n^2)$ complexity arises from smart meters mapping real number energy consumption to predefined range levels and performing budget calculations, which involve addition, subtraction, and multiplication in iterative loops for all appliances over time.
Other benchmarking approaches have a complexity of $O(n)$ in the noise addition scenario. Despite our approach having slightly higher computational complexity, we achieve efficiency by handling all appliances simultaneously during noise addition and communicating the result only once. In contrast, other approaches require separate calculation and communication for each appliance.

Our approach demonstrated excellent utility (i.e., hit rate) performance compared to other benchmarking methods, both in terms of average and range. It excels in utility performance by efficiently managing disaggregated energy consumption data, and preserving individual appliance-level details with strong privacy guarantees.

% Our scheme outperforms the benchmarking approaches in terms of utility and time complexity and provides additional benefits, as discussed in Table \ref{tab:compare}. 

%In the absence of any existing methods for direct comparison in top-k appliance sharing, 
We further qualitatively compared our results to existing privacy-preserving approaches for the aggregated value of smart meter data sharing, as presented in Table \ref{tab:compare}. Compared to existing works, our approach takes a more granular perspective on sharing disaggregated smart meter energy consumption data. It enables privacy-preserving sharing of individual appliance energy data without revealing the appliance list or energy consumption, significantly improving over existing methods. Furthermore, we address privacy leakages caused by temporal correlation due to repeated data sharing by using adaptive privacy budget methods. However, the limited privacy budget for many appliances and a long time window can impact the utility of our approach. To improve the utility, we had to relax the privacy budget. However, this is not a significant issue in real-world scenarios, as energy service providers with millions of users can still achieve a good level of utility even with a constrained privacy budget.

\begin{table}[ht]
\caption{Qualitative comparison of {\mysys} against the existing methods (\textbf{RR}: Randomized Response).}
\label{tab:compare}
\centering
\resizebox{0.5\textwidth}{!}{
\begin{tabular}{p{1.4cm}|p{1.0cm}|p{1.5cm}|p{1.1cm}|p{2.9cm}}
\hline

\textbf{Method}   & \textbf{Privacy model} & \textbf{Application} & \textbf{Stream processing} & \textbf{Limitation} \\ \hline

Hassan, et al.\cite{hassan2019differential} & Laplace mechanism  & Real-time load monitoring  & No & Fixed time interval perturbation, Can exploit temporal correlation, Limited to total energy consumption \\ \hline
Gai et al. \cite{gai2022efficient}   & RR      & Aggregation and average & No &  Fixed time interval randomization, Can exploit temporal correlation,  Limited to total energy consumption     \\ \hline
\textit{LDP-SmartEnergy} & RR & Top-k appliance energy consumption   & Yes &  Utility reduction by handling privacy in many appliances and repeated data sharing over time with limited privacy budget \\ \hline
\end{tabular}
}
\end{table}

\section{Related Work}
\label{sec:related}

%\subsection{Privacy-Preserving Approaches for Smart Meter Data}
In this section, we discuss areas closely related to privacy-preserving smart meter data sharing approaches and differentially private methods on streaming data.

\subsection{Privacy-Preserving Approaches for Smart Meter Data Sharing}
The literature shows various privacy-preserving approaches for smart meter data, including cryptographic approaches, privacy with demand shaping and load scheduling, and statistical approaches\cite{marks2021differential}.
Cryptographic techniques, such as homomorphic encryption (HE), are employed to secure smart meter data by encrypting the data to prevent unauthorized access or interception. HE, specifically, enables mathematical operations to be performed on encrypted data, producing ciphertext that can be decrypted to yield the result while maintaining the security and privacy of data. 
% However, scalability and low efficiency have been significant challenges for HE. 
Another method for smart meter aggregation is secure multiparty computation (MPC) \cite{danezis2013smart}. MPC allows secure evaluation of a function on private data distributed among untrusted parties. One approach to MPC  is secret sharing \cite{danezis2013smart}, which involves dividing a secret into multiple shares and distributing them among several parties who do not trust each other \cite{10.1145/3133956.3133982}. Two fundamental issues in MPC are the reliance on trusted parties and the need for a significant number of communications \cite{10.1145/3133956.3133982}.
Some other cryptographic methods, such as symmetric \cite{acs2011have} and asymmetric DC-Nets \cite{de2016privacy}, use key sharing but can be problematic for large-scale smart grids. Besides, cryptographic methods are computationally complex and vulnerable to Sybil attacks and have increased communication costs in resource-constrained environments \cite{gai2022efficient, bell2020secure}. Moreover, in the smart meter aggregation protocol, cryptographic approaches often assume that the parties involved are semi-honest, meaning they conduct computations honestly but are curious to learn as much information as possible from the collected data \cite{8552366}. Another issue with these approaches is that computation overhead brought by the encryption functions makes it impractical for smart meters with limited computing resources to conduct frequent encryption, posing a significant challenge to implementing the approach.

Another privacy-preserving approach specific to smart meters is battery load balancing by charging/discharging energy storage (external batteries), which can hide energy consumption time and appliance load signatures \cite{varodayan2011smart, backes2013differentially}. However, it also has issues such as short battery lifespan, cost of installing large batteries, and environmental impact \cite{hassan2019differential}.

Common statistical methods for protecting smart meter privacy include k-anonymity, aggregation, and differential privacy (DP) \cite{marks2021differential}. DP is preferred as it reduces computational overhead while providing privacy protection. 
Previous approaches for implementing DP in smart meter data have used various types of noise, such as Gamma \cite{acs2011have, eibl2017differential}, Laplace \cite{ hassan2019differential, hossain2021cost}, and geometric \cite{hassan2020performance}, with Laplace being the most common. Eibl and Enger \cite{eibl2017differential} introduced point-wise DP, a real-time perturbation of Laplace-based DP, and Hossain \cite{hossain2021cost} proposed a cost-effective DP strategy using a Multi-Armed Bandit algorithm for both static and dynamic reporting of smart meter data. Gai et al. \cite{gai2022efficient} proposed a scheme using randomized response to estimate total or average power consumption while preserving LDP. 
% They applied a discretization algorithm based on conditional probability to reduce the difference between the aggregation and real data aggregation results. 

However, existing approaches primarily focus on aggregated statistics of energy consumption data (e.g., for calculating total or average consumption), often geared towards protecting them for billing and operational purposes, rather than harnessing its potential for value-added services such as identifying the impact of individual appliances on overall energy consumption, segmenting consumers based on energy usage patterns, or detecting anomalies \cite{asghar2017smart}. However, implementing these value-added services necessitates more granular data, specifically disaggregated energy consumption data of individual appliances, which constitutes the primary focus of this paper. Existing approaches cannot be directly applied to the disaggregated data obtained through ED as they could disclose the individual appliance data (appliance list and appliance's energy value) due to insufficient randomization. Another notable gap is the consideration of smart meter data as a form of streaming data, necessitating time-window processing to mitigate the risk of gradual information disclosure, an aspect that existing LDP methods do not incorporate.

\subsection{Differential Privacy on Data Streams}

Dwork et al. \cite{dwork2010differential} proposed an event-level DP algorithm based on a binary tree technique for finite streams that hides a single event of a user. This approach was developed further in subsequent works \cite{chan2011private, chen2017pegasus}. However, event-level privacy is insufficient for protecting users' privacy over stream data. Hence, user-level DP on finite streams \cite{fan2013adaptive} and w-event DP \cite{kellaris2014differentially} on infinite streams were proposed.

Erlingsson et al. \cite{erlingsson2014rappor} introduced a memoization mechanism over a data stream in cases where the underlying true value changes in an uncorrelated manner. The authors in \cite{kim2018privacy, arcolezi2021longitudinal} used various techniques to prevent average attacks and correlated or non-correlated events in longitudinal analysis; however, only for finite streams. Event-level \cite{wang2021continuous} and user-level LDP \cite{bao2021cgm} are also adopted in literature; however, they can only be applied to finite streams. Ren et al. \cite{10.1145/3514221.3526190} extended work presented in \cite{kellaris2014differentially} for an LDP environment with infinite stream data. They proposed a population division-based approach that utilizes subsampling of users to preserve DP under parallel composition.
Our approach builds upon this work for handling disaggregated energy consumption stream data, as opposed to previous methods, which have utilized the corresponding technique primarily for frequency estimation of categorical data.

\vspace{-1mm}

\section{Conclusion}
\label{sec:conclusion}

We proposed a local differentially private (LDP) approach, named {\mysys}, for sharing disaggregated energy consumption data of household appliances from smart meters over time. {\mysys} employs randomized response and optimized unary encoding techniques to discretize and randomize the energy consumption data, thus ensuring LDP. Hence, {\mysys} tackles the challenges related to untrusted users and servers by employing a unique LDP modeling technique for processing disaggregated energy data. To accommodate the streaming nature of this data, we integrated a sliding window technique into {\mysys}.

We demonstrated the effectiveness of our solution through a thorough data analysis (utilizing a comprehensive use-case) where top-k appliance usage patterns and the appliances list are obtained from energy consumption data from a household. The results of our analyses on {\mysys} demonstrate its capability to accurately release top-k appliance smart meter energy consumption data while maintaining a high level of privacy (e.g., $\varepsilon$ = 10, which is a constrained configuration for data streams) in scenarios featuring 15 appliances. Moreover, our approach achieved a hit rate of 8 out of 10 (5.7 out of 10 on average) in predicting the top 10 appliances, demonstrating a high accuracy. Our approach successfully maintains the privacy and utility of individual users by preserving consistent patterns of aggregated data, even against small datasets and strict privacy constraints.
% Overall, this method shows promise for practical applications in various fields, such as network usage patterns of devices.
Despite its potential, {\mysys} presents several limitations. 
Hence, our future work will address these limitations such as the need for more users to improve utility, generalizability challenges in diverse smart home environments, and the impact of long-term data sharing on utility. 

\section*{Acknowledgement}
The work has been supported by the Cyber Security Research Centre Limited whose activities are partially funded by the Australian Government’s Cooperative Research Centres Program.


\bibliographystyle{ieeetr}
\bibliography{template}

\appendix

\section{Appendices}
\label{appendix}

\subsection{Adaptive Budget Division Methods: LBD and LBA}
\label{app:adaptive_lbd_lba}

\begin{figure}[h!tb]
\centering
  \includegraphics[width=\linewidth]{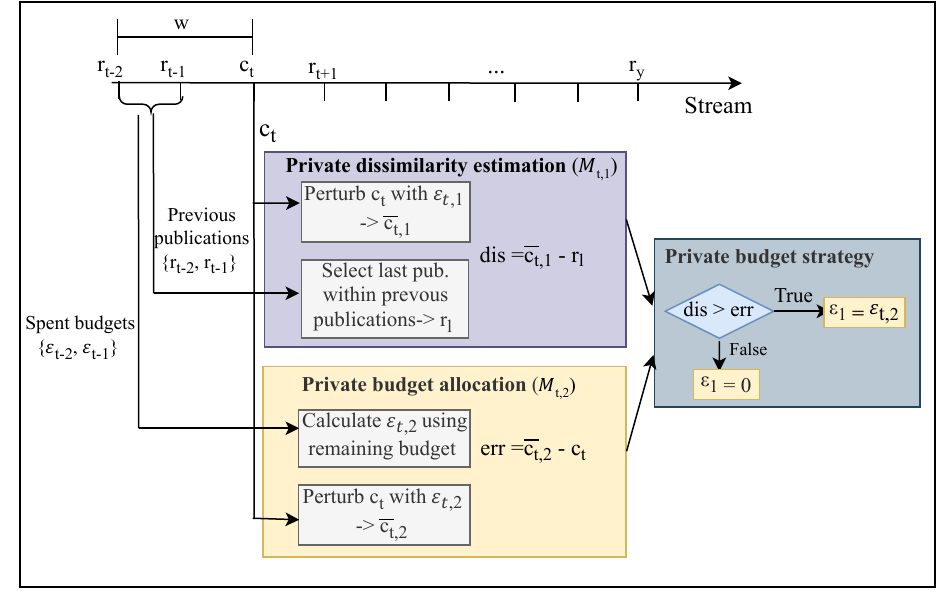}
  \caption{Overview of Adaptive Privacy Budget Division on Streaming Data using LBD and LBA methods. 
  \textbf{$dis$}: The difference between the previous release ($r_l$) and the current perturbed release ($\bar{c}_t$), \textbf{$err$}: The difference between the current perturbed data $\bar c_t$ and the current true data $c_t$, $\varepsilon_1$: The privacy budget calculated for the current timestamp for the randomization of current data. }
  
  \label{fig:workflowadaptive}
\end{figure}

This subsection provides a detailed explanation of the adaptive budget division methods: LBD and LBA. The main steps of LBD and LBA are illustrated in Figure \ref{fig:workflowadaptive}, which helps to clarify the process.

Algorithm \ref{algo:lbdlba} outlines the steps of the adaptive budget division methods, namely the LBD and LBA algorithms. The corresponding notations and their descriptions are presented in %the first part of 
Table \ref{tab:TableOfNotations}. 
% The latter part of the table describes the notations used in Algorithm \ref{alg1}.

\begin{table}[]
\caption{Table of notations for algorithms}
\centering 
\resizebox{0.5\textwidth}{!}{
\begin{tabular}{p{1.8cm}|p{7cm}}
\hline

\textbf{Notations}  & \textbf{Description} \\ \hline
% \multicolumn{2}{p{0.9\linewidth}}{ \centering\textbf{Notations for Algorithm \ref{algo:lbdlba}}} \\ \hline
$M$ & An LDP mechanism processes an input stream and produces DP data as output \\ \hline
$M_{t,1}$, $M_{t,2}$, $M_{t,3}$ & The mechanism $M$ is decomposed into sub-mechanisms such as $M{t,1}$, $M_{t,2}$, $M_{t,3}$, each of which generates independent randomness and collectively achieve $\varepsilon$-LDP \\ \hline
$w$ & A window of up to $w$ timestamps \\ \hline
$\varepsilon$ & Entire privacy budget for $w$ timestamps \\ \hline
\{$r_1$,$r_2$, $\dots$, $r_{l}$\} & Set of previous releases \\ \hline
$c_t$ & True data of current timestamp \\ \hline
$l$ & Lastly published timestamp \\ \hline
$r_{l}$ & The lastly published release with perturbation \\ \hline
$\varepsilon_t,_1$ & Dissimilarity budget allocated in $M{t,1}$ \\ \hline
$\bar c_{t,1}$ & Perturbed $c_t$ using $\varepsilon_{t,1}$  \\ \hline
$dis$ & the dissimilarity measure as the error between the $\bar c_t$ and the $r_{l}$ \\ \hline
$\varepsilon_{rm}$ & The remaining publication budget at current timestamp \\ \hline
$\varepsilon_t,_2$ & The potential publication budget allocated at the current timestamp $t$, which is used to calculate the potential error in $M_{t,2}$. \\ \hline
$\bar c_{t,2}$ & Perturbed $c_t$ using $\varepsilon_{t,2}$  \\ \hline
$err$ & The error / dissimilarity measure between $\bar c_{t,2}$ and $c_t$\\ \hline
$r_t$ & Releasing perturbed data for the current timestamp $t$ \\ \hline
$\varepsilon_{l,2}$ & Used publication budget at the publication timestamp $l$. This doesn't need to be the previous timestamp (t-1), as some timestamps can be skipped. \\ \hline
$t_N$ & Timestamps to be nullified \\ \hline
$t_A$ & Timestamps that can be absorbed \\ \hline
% \multicolumn{2}{p{0.9\linewidth}}{ \centering\textbf{Notations for Algorithm \ref{alg1}}} \\ \hline
% $n$ & Number of appliances \\ \hline
% \{<$a_1$,$v_1$>, $\dots$, <$a_n$, $v_n$>\} &  $n$ Number of appliances and their corresponding energy values \\ \hline
%  \{(0,$r_1$):$l_1$, $\dots$, ($r_d$, max):$l_d$\} & Energy ranges \\ \hline
%  $d$ & Number of bits in the binary array, which represents the total number of energy levels \\ \hline
\multicolumn{2}{p{0.9\linewidth}}{* Here, the second subscript (e.g., $c_{t,2}$) specifies the sub-mechanism number (1, 2, 3). } \\
\end{tabular}
}
\label{tab:TableOfNotations}
\end{table}

\label{algorithm:adaptive}

\setlength{\textfloatsep}{0.5cm}% Remove \textfloatsep
\begin{algorithm*}[h]
\caption{The algorithm for adaptive budget division methods: LBD and LBA}
\label{algo:lbdlba}
\begin{multicols}{2}
         
    \KwIn{Total privacy budget $\varepsilon$, window size $w$, current $c_t$, previous releases \{$r_1$,$r_2$, $\dots$, $r_{l}$\}}
        \KwOut{Released perturbed data }
    
    \textit{Approach 1: Local budget distribution} 

    Initialize $c_t$ = $<0,\dots, 0>^{d}$;
    
    \For {each timestamp $t$} {

    \tcp{\footnotesize Sub Mechanism $M_{t,1}$}
    
    Set $\varepsilon_t,_1$ = $\varepsilon/(2w)$; \label{step:e1}
    
    Perturb $c_t$ using $\varepsilon_t,_1$ $\rightarrow$ $\bar c_{t,1}$
    \label{step:dis2}
    
    Select lastly published release $r_{l}$ from \{$r_1$,$r_2$, $\dots$, $r_{l}$\}
    
    Calculate \textit{dis} = $\frac{1}{d} \sum_{k=1}^{d}(\bar c_{t,1}[k]-r_{l}[k])^{2}$ \label{step:discal}

    \tcp{\footnotesize Sub Mechanism $M_{t,2}$}
    Calculate $\varepsilon_{rm}$ =  $\varepsilon/2$ - $\sum_{i=t-w+1}^{t-1}$$\varepsilon_i,_2$; \label{step:m2}
    
    Set $\varepsilon_t,_2$ = $\varepsilon_{rm}$/2; \label{step:e2}

    Perturb $c_t$ using $\varepsilon_{t,2}$ $\rightarrow$ $\bar c_{t,2}$ \label{step:m23}
    
    Calculate \textit{err} = $\frac{1}{d} \sum_{k=1}^{d}(\overline{c}_{t,2}[k]]-c_{t}[k])^{2}$; \label{step:m2error}

    \tcp{\footnotesize Sub Mechanism $M_{t,3}$}
     \uIf{$dis > err$}{\label{step:check1}
        \tcp{\footnotesize Publication strategy}
        \Return $r_t$ = 
        $\bar c_t,_2$  (perturbed using $\varepsilon_{t,2}$)\label{step:check2}
    }
     \Else{
        \tcp{\footnotesize Approximation strategy}
        \Return $r_t$ = $r_{l}$; set
         $\varepsilon_t,_2$ = 0 \label{step:check3}
     }     
    }

    \textit{Approach 2: Local budget absorption}

    Initialize $c_t$ = $<0,\dots, 0>^{d}$; last publication timestamp $l$ = 0, and $\varepsilon_{l,2}$ = 0;
    
     \For {each timestamp t} {

     \tcp{\footnotesize Sub Mechanism $M_{t,1}$}
        Same as lines \ref{step:e1} to \ref{step:discal}
        
     \tcp{\footnotesize Sub Mechanism $M_{t,2}$}
     
        Calculate $t_N$ = $\frac{\varepsilon_{l,2}}{\varepsilon/(2w)}-1$ \label{step:lbam21}
    
     \uIf{$t-l \leq t_N$}{
        \Return $r_t$ = $r_{l}$; set
        $\varepsilon_t,_2$ = 0     \label{step:null}
     }

     \Else{
        Calculate $t_A$ = $t -(l +t_N$);\label{step:abs1}
        
        Set $\varepsilon_{t,2}$ = 
        $\varepsilon$/(2$w$) $\times$ min($t_A$,$w$); 

        Randomize the current data $c_t$ with $\varepsilon_t,_2$ $\rightarrow$ $\bar c_t,_2$ \label{step:abs2}
        
        Calculate \textit{err} = $\frac{1}{d} \sum_{k=1}^{d}(\overline{c}_{t,2}[k]]-c_{t}[k])^{2}$ \label{step:calerr2}
    
     \tcp{\footnotesize Sub Mechanism $M_{t,3}$}
     \uIf{$dis > err$}{\label{step:check12}
        \tcp{\footnotesize Publication strategy}
        \Return $r_t$ = 
        $\bar c_t,_2$  (perturbed using $\varepsilon_{t,2}$); set $l$ = $t$
        }
         \Else{
            \tcp{\footnotesize Approximation strategy}
            \Return $r_t$ = $r_{l}$; set
             $\varepsilon_t,_2$ = 0 \label{step:check22}
         }     
     }
     }
     \end{multicols}
\end{algorithm*}

\newpage
\subsection{Privacy analysis: LBD and LBA}
\label{subsec:appLBD}

\textbf{(1) LBD satisfies $w$-event LDP}
\begin{proof}
\label{proof:lbd}
LBD comprises two sub-mechanisms ($M_1$ and $M_2$) applied in sequence.

\textit{Sub-mechanism $M_1$}: 

Dissimilarity budget $\varepsilon_{t,1}$ at each timestamp $t$  = $\varepsilon/(2\times w)$. Then, for every $t$ within a $w$ sized window,

\begin{equation}
    \label{eq:subm1}
    \sum_{k=t-w+1}^{t} \varepsilon_{k,1} = \varepsilon/2\omega \times \omega = \varepsilon/2
\end{equation}

% The dissimilarity budget $\varepsilon_{t,1}$ at each timestamp $t$ is $\varepsilon/(2\times w)$. Thus, all timestamps within a window that have the maximum privacy budget will be $\varepsilon/2$.

    % \begin{equation}
    %     \label{eq:subm1}
    %     \sum_{k=t-w+1}^{t} \varepsilon_{k,1} = \varepsilon/2\omega \times \omega = \varepsilon/2
    % \end{equation}

\textit{Sub-mechanism $M_2$}: 

If publication occurs, the maximum privacy budget allocation at each timestamp $t$ within $w$ is $\varepsilon_{t,2} = \varepsilon_{rm}/2 = (\varepsilon/2 -\sum_{k=t-w+1}^{t} \varepsilon_{k,2})/2$ .

Scenario 1: For $1 \le t \le w$, as LBD distributes the budget in a sequence of exponential pattern  $\varepsilon$/4, $\varepsilon$/8, $\dots$, resulting in at most $w$ publications in a time window.
% As LBD follows an exponential sequence pattern in allocating $\varepsilon_{t,2}$, with a maximum of $w$ publications. 
Thus, it follows Eq. (\ref{eq:subm2_1}).

\begin{equation}
    \label{eq:subm2_1}
    \sum_{i=1}^{t}\varepsilon_{k,2}\le (\varepsilon/2).(1-\frac{1}{2^{w}})\le (\varepsilon/2)
\end{equation}

Scenario 2: for $ t \ge w$, $ t = w + m$, then according to Eq. (\ref{eq:subm2_1}),  $\sum_{k=m+1}^{w+m} \varepsilon_{k,2} \le \varepsilon/2$. For next timestamp $t = w + m + 1$, there is,

\begin{equation}
    \label{eq:subm2_2}
    \sum_{k=m+2}^{w+m+1} \varepsilon_{k,2} = \sum_{k=m+2}^{w+m} \varepsilon_{k,2} + \varepsilon_{w+m+1,2}
\end{equation}

However, $\varepsilon_{w+m+1,2}$ is half of the remaining budget at time $w+m+1$, so, 

\begin{equation}
    \label{eq:subm2_3}
    % \begin{aligned}
    \varepsilon_{w+m+1,2} \le \varepsilon_{rm}/2 \le (\varepsilon/2 - \sum_{k=m+2}^{w+m} \varepsilon_{k,2})/2
    % \end{aligned}
\end{equation}

So, when we replace Eq. (\ref{eq:subm2_3}) in Eq. (\ref{eq:subm2_2}),

\begin{equation}
    \label{eq:subm2_4}
    \begin{aligned}     
    \sum_{k=m+2}^{w+m+1} \varepsilon_{k,2} &= \sum_{k=m+2}^{w+m} \varepsilon_{k,2} + (\varepsilon/2 - \sum_{k=m+2}^{w+m} \varepsilon_{k,2})/2 \\
    &= \varepsilon/4 + (\sum_{k=m+2}^{w+m+1} \varepsilon_{k,2})/2 \le \varepsilon/4 + \varepsilon/4 \\
    &\le \varepsilon/2
     \end{aligned}
\end{equation}

For $1 \le t \le w$ and for $t = w + m$, $t = w + m + 1$, if $\sum_{k=t-w+1}^{t}\varepsilon_{k,2} \leq \varepsilon/2$, then it holds for every timestamp $t \geq 1$.

As LBD applies $M_1$ and $M_2$ sequentially at each timestamp $t$, the total privacy budget in a window of size $w$ is the sum of $\varepsilon_{k,1}$ and $\varepsilon_{k,2}$ which is less than or equal to $\varepsilon$.

\end{proof}

\textbf{(2) LBA satisfies $w$-event LDP}
\begin{proof}
\label{proof:lba}

The sub-mechanism $M_1$ in LBA is identical to LBD. So, for each timestamp, $\sum_{k=t-w+1}^{t} \varepsilon_{k,1} = \varepsilon/2\omega \times \omega = \varepsilon/2$.

Three possible scenarios in sub-mechanism $M_2$ is $\varepsilon_2$ may be nullified or be absorbed or absorb unused timestamps from previous timestamps (refer to Section \ref{subsec:lba}).

Scenario 1: the current timestamp $t$ is a publication timestamp that consumes the budget from the previous $\alpha$ timestamps. 

In this case, the publication budgets of the preceding $\alpha$ timestamps ($i\in [t-\alpha, i-1]$) and succeeding $\alpha$ timestamps ($i\in [t+1, t+\alpha]$) are set to 0. Thus, the publication budget for the current timestamp is,

\begin{equation}
    \label{eq:lbasubm2_1}    
    \varepsilon_{t,2} = (1+\alpha)\times \frac{\varepsilon}{2\times w}
\end{equation}

Any window of size $w$ sliding over timestamp $i$ must cover at least $\alpha$ timestamps with $\varepsilon_{i,2}$ = 0.

Suppose there are $n$ timestamps with zero budget due to nullification by timestamp $t$. The sum of the publication budget of timestamp $t$ and the $n$ zero-budget timestamps is at most $(1 +\alpha) \times \frac{\varepsilon}{2w}$, equivalent to each of the $n + 1$ timestamps being assigned with uniform budget of $(1+ \alpha) \times \frac{\varepsilon}{2w(n+1} )\le \frac{\varepsilon}{2w}$. 

This also applies to any other publication timestamp $t'$ that has absorbed its unused budget from the previous timestamps within the same time window as $t$.

So, the total publication budget in a time window $w$, for $M_2$ is,

\begin{equation}
   \sum_{k=t-w+1}^{t} \varepsilon_{k,2} \le \frac{\varepsilon}{2w}\times w = \varepsilon/2
\end{equation}

Similarly LBA applies $M_1$ and $M_2$ sequentially at each timestamp $t$, and the sum of $\varepsilon_{k,1}$ and $\varepsilon_{k,2}$ in a window of size $w$ is $\leq \varepsilon$.

\end{proof}

\subsection{More Experimental Results on {\mysys} Using Synthetic Data}
\label{sec:more_results}

We conducted additional experiments to investigate the impact of the number of users and the number of appliances using the LBD method, which is demonstrated in right-skewed distributions (gives better results than other distributions). 
To assess the utility in relation to the number of users, we conducted experiments where we varied the number of users in the dataset (100, 1000, 10000). As expected, according to LDP theory, when the number of users increases, the two distributions become more similar. This observation is demonstrated in Figure \ref{fig:skewp:userComparison}. 

Figure \ref{fig:skewp:applianceComparison} depicts the impact of the number of appliances on utility. As the number of appliances directly affects the appliance's sensitivity (2 $\times$ $n$), the performance tends to decrease when the number of appliances increases.

\pgfplotsset{compat=newest}
\begin{figure}[!t]
  % \centering
  \begin{subfigure}[b]{0.23\textwidth}
  \begin{tikzpicture}
    \begin{axis}[
        % /pgf/number format/1000 sep={},
        width=0.8\textwidth,
        height=1.2in,
        scale only axis,
        clip=false,
        separate axis lines,
        axis on top,
        xmin=0,
        xmax=4,
        xtick={1,2,3},
        x tick style={draw=none},
        yticklabel style={font=\scriptsize},
        xticklabel style={font=\scriptsize},
        xticklabels={100, 1000, 10000},
        ymin=0,
        y label style={at={(-0.22,0.25)}, anchor=west},
        ylabel style={font=\small},
        ylabel={p-values},
        xlabel style={font=\small},
        xlabel={Number of users},
        every axis plot/.append style={
          ybar,
          bar width=.5,
          bar shift=0pt,
          fill
        }
      ]
      \addplot[coloruser100]coordinates {(1,0.00038)};
      \addplot[coloruser1000]coordinates{(2,0.00088)};
      \addplot[coloruser10000]coordinates{(3,0.00175)};
    \end{axis}
  \end{tikzpicture}
  \caption{Different number of participated users}
  \label{fig:skewp:userComparison}
  \end{subfigure}
  \begin{subfigure}[b]{0.23\textwidth}
  \begin{tikzpicture}
    \begin{axis}[
        % /pgf/number format/1000 sep={},
        width=0.82\textwidth,
        height=1.2in,
        scale only axis,
        clip=false,
        separate axis lines,
        axis on top,
        xmin=0,
        xmax=5,
        xtick={1,2,3,4},
        x tick style={draw=none},
        yticklabel style={font=\scriptsize},
        xticklabel style={font=\scriptsize},
        xticklabels={5, 10, 15, 20},
        ymin=0,
        y label style={at={(-0.15,0.25)}, anchor=west},
        ylabel style={font=\small},
        ylabel={p-values},
        xlabel style={font=\small},
        xlabel={Number of appliances},
        every axis plot/.append style={
          ybar,
          bar width=.5,
          bar shift=0pt,
          fill
        }
      ]
      \addplot[colora5]coordinates {(1,0.00025)};
      \addplot[colora10]coordinates{(2,0.00018)};
      \addplot[colora15]coordinates{(3,0.00016)};
      \addplot[colora20]coordinates{(4,0.00015)};
    \end{axis}
  \end{tikzpicture}
  \caption{Different number of appliances} 
  \label{fig:skewp:applianceComparison}
  \end{subfigure}
 \caption{The Kruskal--Wallis similarity test p values in different scenarios (in synthetic data)}
 \label{fig:three graphs}
\end{figure}
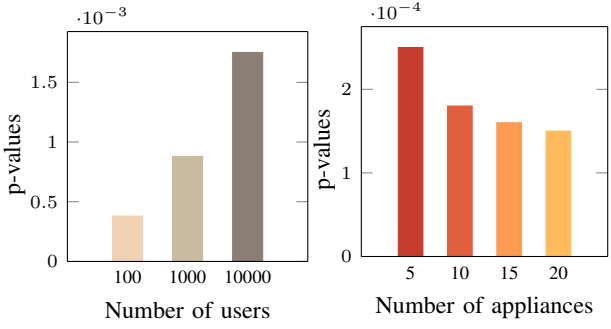

The detailed p-values of each appliance for the discussed scenarios are summarized in Table \ref{tab:data_report}.

\begin{table*}[ht]
\caption{Detailed summary of p-values from all appliances in given scenarios}
\centering
\begin{adjustbox}{angle=90}
\resizebox{1.5\textwidth}{!}{
% \begin{tabularx}{\textwidth}{|X|X|X|X|X|X|X|X|X|X|X|X|X|X|X|X|X|X|}
\begin{tabular}{|l|c|l|l|l|l|l|l|l|l|l|l|l|l|l|l|l|l|}
\hline
\multicolumn{1}{|c|}{Data}                   & \multicolumn{1}{l|}{Scenarios}                        &        & A1                            & A2        & A3        & A4        & A5        & A6        & A7        & A8        & A9        & A10       & A11       & A12       & A13       & A14       & A15       \\ \hline
\multirow{16}{*}{IDEAL Dataset}              & \multirow{4}{*}{Methods}                              & LBU    & 0.000155                      & 0.000156  & 000156    & 0.000156  & 0.000421  & 0.000156  & 0.000155  & 0.000159  & 0.000158  & 0.00151   & 0.000155  & 0.001469  & 0.00015   & 0.001549  & 0.000208  \\ \cline{3-18} 
                                             &                                                       & LSP    & 0.000156                      & 0.000156  & 0.000155  & 0.000156  & 0.00194   & 0.000156  & 0.000159  & 0.00192   & 0.00193   & 0.00202   & 0.00192   & 0.00193   & 0.000156  & 0.00203   & 0.00194   \\ \cline{3-18} 
                                             &                                                       & LBD    & 0.000155                      & 0.000156  & 0.000156  & 0.000155  & 0.00117   & 0.000156  & 0.000155  & 0.000202  & 0.000211  & 0.00181   & 0.000155  & 0.00173   & 0.000155  & 0.00185   & 0.00055   \\ \cline{3-18} 
                                             &                                                       & LBA    & 0.000156                      & 0.000156  & 000155    & 0.000156  & 0.00184   & 0.000156  & 0.000155  & 0.000939  & 0.00103   & 0.00189   & 0.000189  & 0.00176   & 0.000155  & 0.00196   & 0.00168   \\ \cline{2-18} 
                                             & \multicolumn{1}{l|}{\multirow{4}{*}{Window size}}     & w = 2  & 0.000155                      & 0.000156  & 0.000155  & 0.000156  & 0.00189   & 0.000156  & 0.000155  & 0.00174   & 0.00171   & 0.00193   & 0.000527  & 0.00183   & 0.000155  & 0.00197   & 0.00185   \\ \cline{3-18} 
                                             & \multicolumn{1}{l|}{}                                 & w = 3  & 0.000156                      & 0.000156  & 000155    & 0.000156  & 0.00184   & 0.000156  & 0.000155  & 0.000939  & 0.00103   & 0.00189   & 0.000189  & 0.00176   & 0.000155  & 0.00196   & 0.00168   \\ \cline{3-18} 
                                             & \multicolumn{1}{l|}{}                                 & w = 5  & 0.000156                      & 0.000156  & 0.000156  & 0.000155  & 0.000154  & 0.000156  & 0.000155  & 0.000323  & 0.000344  & 0.00185   & 0.000155  & 0.00171   & 0.000155  & 0.00186   & 0.00102   \\ \cline{3-18} 
                                             & \multicolumn{1}{l|}{}                                 & w = 7  & 0.000156                      & 0.000156  & 0.000156  & 0.000155  & 0.001306  & 0.000156  & 0.000155  & 0.000226  & 0.000236  & 0.00182   & 0.000155  & 0.00169   & 0.000155  & 0.00180   & 0.000574  \\ \cline{2-18} 
                                             & \multicolumn{1}{l|}{\multirow{4}{*}{Privacy budgets}} & e = 5  & 0.000156                      & 0.000156  & 0.000156  & 0.000155  & 0.00138   & 0.000156  & 0.000155  & 0.000245  & 0.000278  & 0.00180   & 0.000155  & 0.00167   & 0.000155  & 0.00184   & 0.000727  \\ \cline{3-18} 
                                             & \multicolumn{1}{l|}{}                                 & e = 10 & 0.000156                      & 0.000156  & 0.000155  & 0.000156  & 0.00184   & 0.000156  & 0.000155  & 0.000939  & 0.00103   & 0.00189   & 0.000189  & 0.00176   & 0.000155  & 0.00196   & 0.00168   \\ \cline{3-18} 
                                             & \multicolumn{1}{l|}{}                                 & e = 15 & 0.000156                      & 0.000156  & 0.000156  & 0.000155  & 0.00188   & 0.000156  & 0.000155  & 0.00174   & 0.00174   & 0.00192   & 0.000555  & 0.00186   & 0.000155  & 0.00196   & 0.00184   \\ \cline{3-18} 
                                             & \multicolumn{1}{l|}{}                                 & e = 20 & 0.000156                      & 0.000156  & 0.000155  & 0.000156  & 0.00192   & 0.000156  & 0.000155  & 0.001870  & 0.00186   & 0.00202   & 0.00150   & 0.00190   & 0.000155  & 0.00197   & 0.00188   \\ \cline{2-18} 
                                             & \multicolumn{1}{l|}{\multirow{4}{*}{Levels}}          & l = 5  & 0.00899                       & 0.00900   & 0.00900   & 0.00899   & 0.0759    & 0.0089    & 0.0090    & 0.0729    & 0.0671    & 0.0754    & 0.0655    & 0.0761    & 0.0657    & 0.0780    & 0.0762    \\ \cline{3-18} 
                                             & \multicolumn{1}{l|}{}                                 & l = 7  & 0.00173                       & 0.00173   & 0.00174   & 0.00173   & 0.00174   & 0.00174   & 0.00173   & 0.0170    & 0.0153    & 0.01740   & 0.0128    & 0.0167    & 0.0150    & 0.0175    & 0.0182    \\ \cline{3-18} 
                                             & \multicolumn{1}{l|}{}                                 & l = 10 & 0.000156                      & 0.000156  & 000155    & 0.000156  & 0.00184   & 0.000156  & 0.000155  & 0.000939  & 0.00103   & 0.00189   & 0.000189  & 0.00176   & 0.000155  & 0.00196   & 0.00168   \\ \cline{3-18} 
                                             & \multicolumn{1}{l|}{}                                 & l = 15 & 3.010e-06                     & 3.002e-06 & 3.026e-06 & 2.961e-06 & 3.201e-05 & 3.043e-06 & 3.042e-06 & 1.442e-05 & 1.257e-05 & 3.454e-05 & 3.316e-06 & 2.843e-05 & 2.289e-06 & 3.465e-05 & 2.754e-05 \\ \hline
\multirow{4}{*}{Uniform distribution}        & \multirow{4}{*}{Methods}                              & LBU    & 0.000156                      & 0.000156  & 0.000156  & 0.000156  & 0.000156  & 0.000156  & 0.000156  & 0.000156  & 0.000156  & 0.000156  & 0.000156  & 0.000156  & 0.000156  & 0.000156  & 0.000156  \\ \cline{3-18} 
                                             &                                                       & LSP    & 0.000156                      & 0.000156  & 0.000156  & 0.000156  & 0.000156  & 0.000156  & 0.000156  & 0.000156  & 0.000156  & 0.000156  & 0.000156  & 0.000156  & 0.000156  & 0.000156  & 0.000156  \\ \cline{3-18} 
                                             &                                                       & LBD    & 0.000156                      & 0.000156  & 0.000156  & 0.000156  & 0.000156  & 0.000156  & 0.000156  & 0.000156  & 0.000156  & 0.000156  & 0.000156  & 0.000156  & 0.000156  & 0.000156  & 0.000156  \\ \cline{3-18} 
                                             &                                                       & LBA    & 0.000156                      & 0.000156  & 0.000156  & 0.000156  & 0.000156  & 0.000156  & 0.000156  & 0.000156  & 0.000156  & 0.000156  & 0.000156  & 0.000156  & 0.000156  & 0.000156  & 0.000156  \\ \hline
\multirow{4}{*}{Skewed (left) distribution}  & \multirow{4}{*}{Methods}                              & LBU    & 0.000140                      & 0.000138  & 0.000139  & 0.000138  & 0.000139  & 0.000138  & 0.000138  & 0.000138  & 0.000138  & 0.000139  & 0.000138  & 0.000138  & 0.000138  & 0.000138  & 0.000139  \\ \cline{3-18} 
                                             &                                                       & LSP    & 0.000138                      & 0.000139  & 0.000138  & 0.000139  & 0.000138  & 0.000138  & 0.000138  & 0.000138  & 0.000139  & 0.000138  & 0.000138  & 0.000138  & 0.000138  & 0.000139  & 0.000139  \\ \cline{3-18} 
                                             &                                                       & LBD    & \multicolumn{1}{c|}{0.000238} & 0.000237  & 0.000241  & 0.000234  & 0.000241  & 0.000247  & 0.000236  & 0.000236  & 0.000238  & 0.000238  & 0.000238  & 0.000258  & 0.000238  & 0.000240  & 0.000240  \\ \cline{3-18} 
                                             &                                                       & LBA    & \multicolumn{1}{c|}{0.00014}  & 0.000138  & 0.000138  & 0.000138  & 0.000138  & 0.000138  & 0.000138  & 0.000138  & 0.000138  & 0.000138  & 0.000139  & 0.000138  & 0.000138  & 0.000138  & 0.000139  \\ \hline
\multirow{4}{*}{Normal distribution}         & \multirow{4}{*}{Methods}                              & LBU    & 0.000156                      & 0.000156  & 0.000156  & 0.000156  & 0.000156  & 0.000156  & 0.000156  & 0.000156  & 0.000156  & 0.000156  & 0.000156  & 0.000156  & 0.000156  & 0.000156  & 0.000156  \\ \cline{3-18} 
                                             &                                                       & LSP    & 0.000156                      & 0.000156  & 0.000156  & 0.000156  & 0.000156  & 0.000156  & 0.000156  & 0.000156  & 0.000156  & 0.000156  & 0.000156  & 0.000156  & 0.000156  & 0.000156  & 0.000156  \\ \cline{3-18} 
                                             &                                                       & LBD    & 0.000156                      & 0.000156  & 0.000156  & 0.000156  & 0.000156  & 0.000156  & 0.000156  & 0.000156  & 0.000156  & 0.000156  & 0.000156  & 0.000156  & 0.000156  & 0.000156  & 0.000156  \\ \cline{3-18} 
                                             &                                                       & LBA    & 0.000156                      & 0.000156  & 0.000156  & 0.000156  & 0.000156  & 0.000156  & 0.000156  & 0.000156  & 0.000156  & 0.000156  & 0.000156  & 0.000156  & 0.000156  & 0.000156  & 0.000156  \\ \hline
\multirow{4}{*}{Skewed (right) distribution} & \multirow{4}{*}{Methods}                              & LBU    & 0.000145                      & 0.000145  & 0.000146  & 0.000145  & 0.000145  & 0.000145  & 0.000145  & 0.000145  & 0.000146  & 0.000145  & 0.000145  & 0.000146  & 0.000145  & 0.000145  & 0.000145  \\ \cline{3-18} 
                                             &                                                       & LSP    & 0.000145                      & 0.000145  & 0.000146  & 0.000145  & 0.000145  & 0.000145  & 0.000145  & 0.000145  & 0.000146  & 0.000145  & 0.000145  & 0.000146  & 0.000145  & 0.000145  & 0.000145  \\ \cline{3-18} 
                                             &                                                       & LBD    & 0.000182                      & 0.000516  & 0.000400  & 0.000454  & 0.000453  & 0.000183  & 0.00122   & 0.000523  & 0.000524  & 0.000523  & 0.000514  & 0.000453  & 0.000454  & 0.000514  & 0.000183  \\ \cline{3-18} 
                                             &                                                       & LBA    & \multicolumn{1}{c|}{0.000145} & 0.000145  & 0.000145  & 0.000146  & 0.000144  & 0.000145  & 0.000145  & 0.000145  & 0.000146  & 0.000145  & 0.000145  & 0.000145  & 0.000145  & 0.000145  & 0.000145  \\ \hline

\end{tabular}
}
\end{adjustbox}
\label{tab:data_report}
\end{table*}

\end{document}